\documentclass[twocolumn]{aastex631}

\usepackage{apjfonts}
\usepackage{bm}
\usepackage{amsmath}
\DeclareSymbolFont{cmletters}{OML}{cmm}{m}{it}
\DeclareMathSymbol{v}{\mathalpha}{cmletters}{"76}

\newcommand{\myvec}{\bm}

\newcommand{\hammer}{H-AMR}
\newcommand{\p}{\partial}

\shorttitle{\hammer{}}
\shortauthors{Liska et al.}
\graphicspath{{./}{figures/}}

\begin{document}

\title[\hammer{}]{\hammer{}: A New GPU-accelerated GRMHD Code for Exascale Computing With 3D Adaptive Mesh Refinement and Local Adaptive Time-stepping}

\correspondingauthor{Matthew Liska}
\email{matthew.liska@cfa.harvard.edu}
\author{M.T.P. Liska}
\affiliation{Institute for Theory and Computation, Harvard University, 60 Garden Street, Cambridge, MA 02138, USA}
\author{K. Chatterjee}
\affiliation{Institute for Theory and Computation, Harvard University, 60 Garden Street, Cambridge, MA 02138, USA}
\author{D. Issa}
\affiliation{Center for Interdisciplinary Exploration \& Research in Astrophysics (CIERA), Physics \& Astronomy, Northwestern University, Evanston, IL 60202, USA}
\author{D. Yoon}
\affiliation{Anton Pannekoek Institute for Astronomy (API), University of Amsterdam, Science Park 904, Amsterdam, The Netherlands}
\author{N. Kaaz}
\affiliation{Center for Interdisciplinary Exploration \& Research in Astrophysics (CIERA), Physics \& Astronomy, Northwestern University, Evanston, IL 60202, USA}
\author{A. Tchekhovskoy}
\affiliation{Center for Interdisciplinary Exploration \& Research in Astrophysics (CIERA), Physics \& Astronomy, Northwestern University, Evanston, IL 60202, USA}
\author{D. van Eijnatten}
\affiliation{Anton Pannekoek Institute for Astronomy (API), University of Amsterdam, Science Park 904, Amsterdam, The Netherlands}
\author{G. Musoke}
\affiliation{Anton Pannekoek Institute for Astronomy (API), University of Amsterdam, Science Park 904, Amsterdam, The Netherlands}
\author{C. Hesp}
\affiliation{Anton Pannekoek Institute for Astronomy (API), University of Amsterdam, Science Park 904, Amsterdam, The Netherlands}
\author{V. Rohoza}
\affiliation{Center for Interdisciplinary Exploration \& Research in Astrophysics (CIERA), Physics \& Astronomy, Northwestern University, Evanston, IL 60202, USA}
\author{S. Markoff}
\affiliation{Anton Pannekoek Institute for Astronomy (API), University of Amsterdam, Science Park 904, Amsterdam, The Netherlands}
\author{A. Ingram}
\affiliation{School of Mathematics, Statistics and Physics, Newcastle University, Herschel Building, Newcastle upon Tyne NE1 7RU, United Kingdom}
\author{M. van der Klis}
\affiliation{Anton Pannekoek Institute for Astronomy (API), University of Amsterdam, Science Park 904, Amsterdam, The Netherlands}



\begin{abstract}
  General-relativistic magnetohydrodynamic (GRMHD) simulations have revolutionized our understanding of black hole accretion. Here, we present a graphics processing unit (GPU) accelerated GRMHD code \hammer{} with multi-faceted optimizations that, collectively, accelerate computation by 2-5 orders of magnitude for a wide range of applications. Firstly, it introduces a spherical grid with 3D adaptive mesh refinement that operates in each of the 3 dimensions independently. This allows us to circumvent the Courant condition near the polar singularity, which otherwise cripples high-resolution computational performance. Secondly, we demonstrate that local adaptive time-stepping (LAT) on a logarithmic spherical-polar grid accelerates computation by a factor of $\lesssim10$ compared to traditional hierarchical time-stepping approaches. Jointly, these unique features lead to an effective speed of $\sim10^9$ zone-cycles-per-second-per-node on 5,400 NVIDIA V100 GPUs (i.e., 900 nodes of the OLCF Summit supercomputer). We illustrate \hammer{}'s computational performance by presenting the first GRMHD simulation of a tilted thin accretion disk threaded by a toroidal magnetic field around a rapidly spinning black hole. With an effective resolution of $13$,$440\times4$,$608\times8$,$092$ cells, and a total of $\lesssim22$ billion cells and $\sim0.65\times10^8$ timesteps, it is among the largest astrophysical simulations ever performed. We find that frame-dragging by the black hole tears up the disk into two independently precessing sub-disks. The innermost sub-disk rotation axis intermittently aligns with the black hole spin, demonstrating for the first time that such long-sought alignment is possible in the absence of large-scale poloidal magnetic fields.
\end{abstract}


\section{Introduction}
\label{ch1:sec:introduction}
GRMHD simulations in conjunction with radiative transfer calculations provide arguably the most direct link between the physical laws describing the motion of gas and magnetic fields near black holes (BHs), and the observed phenomenology of accretion disks and jets in astrophysical black hole systems. Initially, the numerical work focused on the geometrically-thick radiatively inefficient accretion flows (RIAFs, e.g., \citealt{Narayan1994, Narayan1995A}), with the accretion typically proceeding in the equatorial plane of a spinning black hole (e.g., \citealt{Villiers2003,hirose04,mck04,mck05,Hawley2006,Beckwith2008,bhk08}).  Geometrically-thin accretion discs
\citep{ss73} are many orders of magnitude more luminous than RIAFs (e.g. \citealt{Sikora2007, Jones2016}) and thought to be responsible for most of the BH growth. Over time, it became possible for the simulations to resolve not only the small thickness of thin disks but also the smaller length scales generated by the magnetized turbulence \citep{Shaffee2008,penna10,Noble2010}, which is powered by the magnetorotational instability (MRI, \citealt{Balbus1991}). 

However, recently, evidence has emerged that much higher resolutions than currently achievable by state-of-the-art simulations (e.g., $N_r\times N_\theta\times N_\varphi = 256^3$ cells, where $N_i$ is the number of cells in the $i$-th direction) are needed to properly resolve disk turbulence in global simulations of magnetized black hole accretion. Namely, a key GRMHD code comparison project \citep{Porth2019} found that even the resolution of $384^3$ cells is insufficient to reach convergence in the critical parameters such as the value of the effective $\alpha$-viscosity generated by the magnetized turbulence. To make matters worse, typically the timescales covered by GRMHD simulations falls short by order(s) of magnitude compared to observationally-interesting timescales, e.g., the timescale to generate and advect large-scale poloidal magnetic fields from large radii \citep{Liska2018B}, to propagate active galactic nuclei (AGN) jets to galaxy scales (e.g \citealt{Mckinney2006, Chatterjee2019, Lalakos2022}), and to evolve tidal disruption events for a fallback time (e.g. \citealt{Shiokawa2015,Curd2019, Andalman2020}).

Observations and theoretical arguments furthermore suggest that accretion systems lack symmetries often taken for granted in GRMHD simulations. For instance, an accretion disk is typically misaligned with respect to the black hole spin axis (e.g. \citealt{Hjellming1995, Greene2001, Volonteri2005, King2005, Caproni2006, Caproni2007}), but this is often not taken into account for expedience.  Such a tilt will, however, significantly change the disk dynamics and the black hole spin evolution (e.g.~\citealt{bp75, Papaloizou1983, Ivanov1997, Ogilvie1999,Lubow2002, Nixon2012B}). In fact, the dragging of inertial frames by the spinning black holes causes tilted disks to precess, which can lead to interesting periodic time variability (e.g.~\citealt{Fragile2005,Fragile2007, Lodato2010, Nixon2012B, Nealon2015}). To understand its potential connection to quasi-periodic oscillations \citep[QPOs,][]{Stella1998, Ingram2009,Ingram2016, Kalamkar2016, Miller2019} observed in the lightcurves of X-ray binaries (XRBs; e.g.~\citealt{Klis1985, Klis2006}), and whether similar variability can be present in AGN, requires even costlier simulations with even longer runtimes. If such simulations were possible, QPOs could possibly be exploited to independently verify black hole spin measurements and provide unique constraints on the disk geometry (e.g.~\citealt{Fragile2007, Nixon2012B, Liska2019b}). 

Indeed, the extremely high resolution and long runtime required to tackle these problems is prohibitive, limiting our ability to understand black hole accretion and its effect, or feedback, on the surrounding ambient medium. The simulation cost scales very steeply with the thickness of the disk, as $(h/r)^{-5}$ with an adaptive mesh [and as $(h/r)^{-6}$ without an adaptive mesh, Sec.~\ref{ch1:sec:Pole}]. For example, to go from a typical thick, $h/r\sim0.3$, low-luminosity accretion disk \citep{Narayan1994} to a typical thin luminous quasar disk that is about an order of magnitude thinner, $h/r\sim0.03$ \citep{ss73}, and misaligned relative to the black hole by a typical angle of 1 radian, would require an order of magnitude increase in resolution in every dimension to fully capture the 3-dimensional structure of MRI turbulence. This would without adaptive and static mesh refinement (Sec.~\ref{sec:amr}), require a factor $\sim 10^3$ more cells and a factor $\sim 30$ more timesteps (Sec.~\ref{ch1:sec:LAT}) to run the simulation for the same amount of time. However, the simulation would need to be run for $\sim 100$ times longer to capture the viscous time, $t_{visc} \propto (h/r)^{-2})$, of the very thin accretion disk \citep{ss73}. All together, a thin, titled disk would increase the cost of a simulation by a factor of $10^5-10^6$, requiring orders of magnitude more computational power than presently feasible. It is therefore one of the most urgent challenges to improve the performance of GRMHD codes and adaptive mesh refinement (AMR) frameworks to bridge this performance gap.

One interesting way of accelerating MHD simulations is to use special hardware such as graphics processing units (GPU). Since GPUs dedicate a much larger fraction of their transistors to raw floating point power compared to central processing units (CPUs), they can be much faster. However, writing efficient code on GPUs is challenging because GPUs require a high level of parallelism and have relatively small caches and low memory bandwidth compared to CPUs. In recent years several groups have utilised GPU-acceleration in astrophysical MHD codes. For example K-ATHENA \citep{Grete2019}, a GPU-accelerated version of Athena++ \citep{Stone2020}, achieves a factor $\sim 10$ speedup on an NVIDIA V100 GPU compared to a 20-core Intel Skylake CPU. The same authors have also demonstrated excellent scale-ability on OLCF Summit, the largest GPU cluster in production at the time of writing. Similar speedups (factor $\sim 10$) were obtained by the GAMER-2 MHD code \citep{Hsi2018}, and the HORIZON \citep{Zink2011} and cuHARM \citep{Begue2022} GRMHD codes. While a factor $\sim 10$ speedup makes existing simulations faster, it is insufficient to attack black hole accretion in the challenging regimes discussed above.

To tackle this problem, we adopt a multi-faceted approach that uses (i) a highly optimized code capable of running both on central processing units (CPUs) and graphical processing units (GPUs), and (ii) a custom-built advanced AMR framework. Namely, we describe various optimizations that we implemented in our new GRMHD code \hammer{} (pronounced ``hammer'') that speed up GRMHD simulations by 2 to 5 orders of magnitude for the especially challenging problems discussed above. In Sec.~\ref{ch1:sec:numerical_methods} we introduce our code and in Sec.~\ref{ch1:sec:Optimizations} we describe our algorithmic advances -- 3D AMR framework and local adaptive timestepping (LAT) -- that provide the bulk of \hammer{}'s algorithmic speedup. In Sec.~\ref{ch1:sec:Optimizations} we describe hardware-specific optimizations on both CPUs and GPUs. In Sec.~\ref{ch1:sec:Scaling} we show benchmarks for various generations of CPUs and GPUs, and in Sec.~\ref{ch1:sec:Scaling} we present the largest ever 3D GRMHD simulation, including the scaling tests, run on OLCF Summit. We summarize our results and provide an outlook for the future development of \hammer{} in Sec.~\ref{ch1:sec:Discussion}.

\newpage
\section{Numerical Scheme}
\label{ch1:sec:numerical_methods}
\hammer{} originally branched out from an open-source HARMPI code \citep{Tchekhovskoy2019}, which derives from publicly available HARM2D code  \citep{Gammie2003,Noble2006}. \hammer{} solves the ideal GRMHD equations in the conservative form. These do not explicitly include any resistive or viscous terms. Instead, \hammer{} relies on numerical dissipation from Riemann solvers to arrive at a physical solution when dissipation is necessary. This is a reasonable approximation since dissipation typically occurs on spatial scales much smaller than what can be resolved numerically in global GRMHD simulations.

Recently, \hammer{} has also incorporated a radiation module (e.g. \citealt{Liska2022}) which uses the M1 approximation (e.g. \citealt{Levermore1984, Mckinney2013}) and can take into account thermal decoupling between ions and electrons (e.g. \citealt{Ressler2015}). We leave the description and validation of our radiation module to a future publication.

\subsection{Riemann Solvers, Spatial Reconstruction and Time Integration}
\hammer{} uses a finite volume shock-capturing Godunov-based HLLE Riemann solver \citep{Harten1983}. To calculate the HLLE fluxes we use an approximation for the fast-wave speed, $v_f$ (e.g. \citealt{Gammie2003}). This overestimates $v_f$ by up to a factor 2 in rare circumstances, which results in additional numerical dissipation, but allows for an analytic solution. Although \hammer{} has the option to solve for the full dispersion relation, we do not describe this feature here since it did not significantly increase the accuracy. In addition to the 2-wave HLLE solver, we have implemented the 3-wave HLLC \citep{Mignone2005, MignoneBodo2006} and 5-wave HLLD \citep{Mignone2009} solvers by transforming the Riemann problem into a locally flat Minkowski frame \citep{Pons1998, Anton2006, White2016}. However, since these less-diffusive solvers can produce numerical artifacts in highly magnetized regions, we will report on their implementation once these issues are fully resolved.

If intercell fluxes are treated as Riemann problems without any spatial reconstruction, this leads to $1^{st}$ order spatial accuracy. However, in smooth flows it is possible to reconstruct the left and right states of the Riemann problem to achieve a higher order of convergence. In \hammer{} we use piecewise parabolic, second order, accurate spatial reconstruction of `primitive' cell variables (PPM, \citealt{Col1984}) on cell faces with an optional shock flattener. Our implementation of PPM does not take into account non-uniform grid spacing (e.g. \citealt{Mignone2014}), but we are considering this for future work. In the GRMHD community-wide code comparison project \citep{Porth2019} we found PPM to be significantly more accurate than other $2^{nd}$ order reconstruction schemes (MINMOD and MC, \citealt{leveque2002}).

The conserved quantities ($U$) are evolved in time by advancing their state from $t^n$ to $t^{n+1/2}$ in a predictor step, and then using the fluxes calculated at $t^{n+1/2}$ to advance the state from $t^n$ to $t^{n+1}$ in a corrector (full) step. The timestep for our midpoint method is limited by the fast wave speed in the coordinate frame through the \citet{Courant1928} condition. We set $\Delta t= C_0 /(1/\Delta t^1+1/\Delta t^2+1/\Delta t^3)$. Here $\Delta t^i=\Delta x^i/v_f$ is the time it takes a fast wave to cross a cell of size $\Delta x^i$ in the $i$-th direction and $C_0$ is the Courant factor that we typically set to $C_0 \sim 0.7-0.9$. Our time integration is similar to the VL2 scheme described in \citet{Stone2009}, but we use spatial reconstruction of variables for both the predictor and corrector steps. This formally does not guarantee that the scheme is total variation diminishing (TVD), but in practice produces more accurate results. In the future, we plan to explore higher order time integration schemes that are strong stability preserving (SSP), including $3^{rd}$ and $4^{th}$ order Runge-Kutta schemes. These might provide additional stability that will be necessary for less diffusive Riemann solvers (e.g. \citealt{Mignone2009}) and higher order spatial reconstruction schemes (e.g \citealt{Tchekhovskoy2007}).

\subsection{Coordinate System and Metric}
Since \hammer{} solves equations of motion in a covariant form, any smoothly varying coordinate system (and metric) can be chosen. In this work the grid is uniform in modified spherical coordinates which are defined as $x^0\equiv t, x^1\equiv\log r, x^2\equiv\theta$, $x^3\equiv\varphi$. This grid extends from an inner spherical radius $R_{\rm in}$, which is located just inside the event horizon, out to an outer radius $R_{\rm out}$. We choose $R_{\rm in}$ such that there are at least $\gtrsim5$ cells between $R_{\rm in}$ and the event horizon and usually set $R_{\rm out} = 10^4 r_g$ or larger. This ensures that both of the radial boundaries do not affect the physical evolution of the system.

The warping of space-time is described in \hammer{} by a metric tensor $g_{\mu \nu}$ with lapse $\alpha=1/\sqrt{-g^{00}}$, shift $\beta^{i}=g^{0i}$, and determinant $g=Det(g_{\mu \nu})$. We use $(\mu, \nu)$ as indices for 4-dimensional and $(i,j)$ as indices for 3-dimensional objects. $\Gamma=-u^{\mu}n_{\mu}=\alpha u^{t}$ is the Lorentz factor corresponding to 4-velocity $u^{\mu}$ as measured by the Zero-Angular-Momentum-Observer (ZAMO) with 4-velocity $n_{\mu}=(-\alpha, 0, 0, 0)$. The spatial projection tensor $\gamma^{\mu}_{\nu}$ is defined as $\gamma^{\mu}_{\nu}=g^{\mu}_{\nu}+n^{\mu}n_{\nu}$ and the 4-dimensional Levi-Civita symbol is $\epsilon^{\mu \nu \lambda \delta}$.

In this work we use a Kerr space-time (unless stated otherwise). We first calculate the Kerr metric tensor analytically in Kerr-Schild coordinates and then transform it numerically to give $g_{\mu \nu}$ in internal coordinates. The corresponding Christoffel symbols, $\Gamma_{\mu \nu}^{\zeta}$, are calculated numerically by taking finite differences of the metric in internal coordinates.

\subsection{Primitive Variables}
We express the conservation laws that \hammer{} evolves (Sec.~\ref{sec:cons_laws}) in terms of fluid frame gas density $\rho$, internal energy density $u_g$, gas pressure $p_g$, fluid-frame magnetic 4-vector $b^{\mu}$, and fluid 4-velocity $u^{\mu}$. These relate to the primitive variables $\rho$, $u_g$, relative 4 velocity $\tilde{u}^{i}$ and coordinate frame magnetic field $B^i$ as follows  (e.g. \citealt{Noble2006}):

 \begin{equation}
 b^{i}=\frac{B^{i}+ b^t u^{i} }{u^{0}}
 \end{equation} 
 
  \begin{equation}
 b^{t}=g_{i \mu}B^{i}u^{\mu}
 \end{equation} 
 
 \begin{equation}
 \tilde{u}^{i}=\gamma^{i}_{\mu}u^{\mu}=u^{i}+\frac{\beta^{i}}{\alpha}\Gamma
 \end{equation}

The relative 4-velocity $\tilde{u^i}$ is numerically convenient because it guarantees that $u^{\mu}u_{\mu}=-1$ for any value of $\tilde{u^i}$. In \hammer{} we store the primitive variables at cell centers. In this manuscript we adopt an ideal gas equation of state with $p_g=(\gamma-1)u_g$ and Lorentz-Heaviside units, $G=M=c=1$. The stress energy tensor $T^{\mu}_{\nu}$, Faraday tensor $\mathcal{F}^{\mu \nu}$, dual of the Faraday tensor $\mathcal{F}^{* \mu \nu}$, and entropy tracer $\kappa$ are given by:

 \begin{equation}
T^{\mu}_{\nu}=(\rho+u_g+p_g)u^{\mu}u_{\nu}+g^{\mu}_{\nu}(p_g+\frac{1}{2}b^{2})-b^{\mu}b_{\nu}
 \end{equation}
  \begin{equation}
\mathcal{F}^{\mu \nu}=-\sqrt{-g}\epsilon^{\mu \nu \lambda \delta}u_{\lambda}b_{\delta}
 \end{equation}
 \begin{equation}
\mathcal{F}^{* \mu \nu}=b^{\mu}u^{\nu}-b^{\nu}u^{\mu}
 \end{equation}
 \begin{equation}
 \kappa=\frac{p_g}{\rho^{\gamma}}
 \end{equation}

The coordinate frame magnetic field $B^i$ and electric field $E^i$ are related to the (dual of the) Faraday tensor as follows:
 \begin{equation}
 B^{i}=\mathcal{F}^{* i t}
 \end{equation}
 \begin{equation}
 E^{i}=\mathcal{F}^{i t}
 \end{equation}

\subsection{Conservation laws}
\label{sec:cons_laws}
\hammer{} evolves the mass (eq.~\ref{eq:mass}), energy (eq.~\ref{eq:ener}), momentum (eq.~\ref{eq:mom}), induction (eq.~\ref{eq:induct}), and entropy (eq.~\ref{eq:entr}) equations. The derivation of these equations from the relativistic Boltzman equation can be found in standard textbooks including \citet{Misner1973} and \citet{RezzollaZanotti2013}. Here we express these equations in covariant form where $\nabla_{\mu}$ denotes a covariant derivative:
  \begin{equation}
 \nabla_{\mu} (\rho u^{\mu})=0
 \label{eq:mass}
 \end{equation}
  \begin{equation}
 \nabla_{\mu} (T^{\mu}_{t})=0
  \label{eq:ener}
 \end{equation}
 \begin{equation}
 \nabla_{\mu} (T^{\mu}_{i})=0
  \label{eq:mom}
 \end{equation}
  \begin{equation}
 \nabla_{\mu} (\mathcal{F}^{* \mu \nu})=0
  \label{eq:induct}
 \end{equation}
  \begin{equation}
 \nabla_{\mu} (\kappa \rho u^{\mu})=0
  \label{eq:entr}
 \end{equation}
For numerical reasons, we subtract the mass conservation equation from the energy equation, i.e. \hammer{} actually evolves $\nabla_{\mu} (T^{\mu}_{t}+\rho u^{\mu})=0$ instead of equation \ref{eq:ener}.

\subsection{Discretisation}
Figure \ref{fig:cell} shows how we evolve the conservation laws (eq.~\ref{eq:mass}-\ref{eq:entr}) on a numerical grid with cell indices $(i,j,k)$ and uniform cell sizes of $(\Delta x^1, \Delta x^2, \Delta x^3)$. We absorb a geometric factor, $\sqrt{-g}$, into the definitions for the volume integrated conserved state $\bar{U}=U\sqrt{-g}$, volume integrated source terms $\bar{Q}=Q\sqrt{-g}$, and area integrated fluxes in the $j$-th direction $\bar{F^j}=F^{j}\sqrt{-g}$. This factor is calculated at the center of each cell for cell-centered quantities and at the center of each face for face-centered quantities in internal coordinates. This is an easy and fast method but slightly less accurate (albeit still $2^{nd}$ order) than explicitly integrating the volume and area over each cell to account for non-uniform grid spacing (e.g. \citealt{Porth2016}).

In \hammer{}, the vector $U=(\rho u^t, T^{t}_{t}, T^{t}_{i}, B^{i}, \kappa \rho u^{t})$ is calculated directly at cell centers from the primitive variables, while $F^j=(\rho u^j, T^{j}_{t}+\rho u^{t}, T^{j}_{i}, b^{j}u^{i}-b^{i}u^{j}, \kappa \rho u^{j})$ is calculated at cell faces from the reconstructed primitive variables using a Riemann solver. The equations of motions take the conservative form, $\frac{\partial U}{\partial t} + \frac{\partial F^j}{\partial x^j} = Q $. With these definitions \textbf{we can write down} the equations of motion in conservative form for the full time step:

\begin{figure}
\begin{center}
\includegraphics[width=\columnwidth,trim=0mm 0mm 0mm 0,clip]{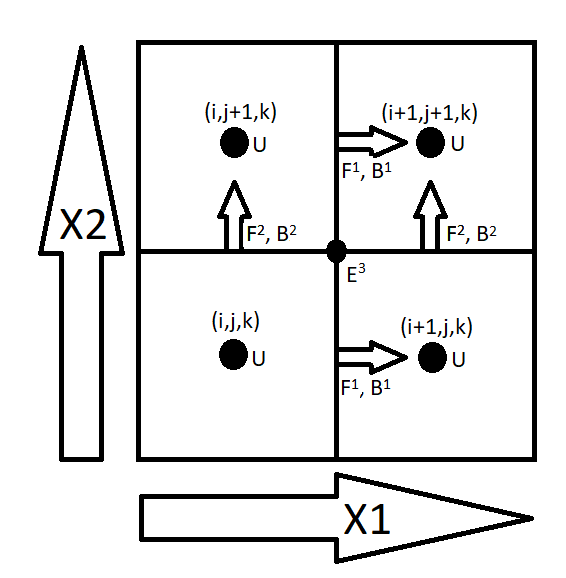}
\caption{An illustration of a cell in the $(x^1, x^2)$ plane in \hammer{}. The conserved quantities $U$ are stored at cell centers, the fluxes $F^{1,2}$ and staggered magnetic field components $B^{1,2}$ are stored at cell faces and the electric fields $E^3$ are stored at cell edges. The corresponding metric components $g_{\mu \nu}$ are stored for cell centers, cell faces, and cell corners. The conserved states are calculated directly from the primitive variables. The fluxes are calculated from the reconstructed primitive variables in a Riemann solver. The electric fields are calculated (in their most simple form) by taking an average of the neighbouring fluxes. Typically, \hammer{} modifies the calculation of the electric fields with a velocity upwinding routine as described in \citet{Gardiner2005}.
}
\label{fig:cell}
\end{center}
\end{figure}

\begin{multline}
  \frac{(\bar{U}^{n+1}_{i,j,k} -\bar{U}^{n}_{i,j,k})}{\Delta t}  = \\
 -\frac{(\bar{F^1})^{n+1/2}_{i+1/2,j,k} -(\bar{F^1})^{n+1/2}_{i-1/2,j,k}}{\Delta x^1} 
  -\frac{(\bar{F^2})^{n+1/2}_{i,j+1/2,k} - (\bar{F^2})^{n+1/2}_{i,j-1/2,k}}{\Delta x^2}\\
  -\frac{(\bar{F^3})^{n+1/2}_{i,j,k+1/2} - (\bar{F^3})^{n+1/2}_{i,j,k-1/2}}{\Delta x^3}+
  \bar{Q}^{n+1/2}_{i,j,k}
\end{multline}
 In the simplest case, the source term $\myvec \bar{Q}$ accounts for the warping of the space-time and grid (e.g. $\bar{Q}_{i}=\Gamma_{\nu i}^{\zeta} T_{\zeta}^{\nu} \sqrt{-g}$ for the momentum equation), but it may also include physical processes such as nuclear heating, gas-radiation interactions, and Coulomb coupling (e.g. \citealt{Liska2022}).

According to Stoke's theorem, to guarantee the divergence-free evolution of the magnetic field the induction equation is evolved on a staggered mesh (e.g. \citealt{EvansHawley1988}). Here, the face-centered magnetic field is advanced in time by taking finite differences in space of the edge-centered electric fields, as we show in Fig.~\ref{fig:cell}: 

\begin{multline}
 \frac{(\bar{B}^1)^{n+1}_{i-1/2,j,k} -(\bar{B}^1)^{n}_{i-1/2,j,k}}{\Delta t}= \\
 \frac{(\bar{E}^2)^{n+1/2}_{i-1/2,j,k+1/2} -  (\bar{E}^2)^{n+1/2}_{i-1/2,j,k-1/2}}{\Delta x^3}-\\
  \frac{(\bar{E}^3)^{n+1/2}_{i-1/2,j+1/2,k} -  (\bar{E}^3)^{n+1/2}_{i-1/2,j-1/2,k}}{\Delta x^2}
\end{multline}
  \begin{multline}
 \frac{(\bar{B}^2)^{n+1}_{i,j-1/2,k} -(\bar{B}^2)^{n}_{i,j-1/2,k}}{\Delta t}=\\
 \frac{(\bar{E}^3)^{n+1/2}_{i+1/2,j-1/2,k} -  (\bar{E}^3)^{n+1/2}_{i-1/2,j-1/2,k}}{\Delta x^1}-\\
  \frac{(\bar{E}^1)^{n+1/2}_{i,j-1/2,k+1/2} -  (\bar{E}^1)^{n+1/2}_{i,j-1/2,k-1/2}}{\Delta x^3}
\end{multline}
 \begin{multline}
 \frac{(\bar{B}^3)^{n+1}_{i,j,k-1/2} -(\bar{B}^3)^{n}_{i,j,k-1/2}}{\Delta t}=\\
 \frac{(\bar{E}^1)^{n+1/2}_{i,j+1/2,k-1/2} - (\bar{E}^1)^{n+1/2}_{i,j-1/2,k-1/2}}{\Delta x^2}-\\
  \frac{(\bar{E}^2)^{n+1/2}_{i+1/2,j,k-1/2} - (\bar{E}^2)^{n+1/2}_{i-1/2,j,k-1/2}}{\Delta x^1}
\end{multline}

We approximate the edge centered electric field by taking the average of the face-centered HLL fluxes $(\bar{F}^{j}(B^{i}) = F^{* j i}\sqrt{-g})$ over $4$ neighbouring cells :

 \begin{multline}
 (\bar{E}^1)_{i,j-1/2,k-1/2}=  0.25[\bar{F}^3(B^{2})_{i,j,k-1/2}+
 \bar{F}^3(B^{2})_{i,j-1,k-1/2}-\\
 \bar{F}^2(B^{3})_{i,j-1/2,k-1}- \bar{F}^2(B^{3})_{i,j-1/2,k}]
\end{multline}
 \begin{multline}
 (\bar{E}^2)_{i-1/2,j,k-1/2}=  0.25[\bar{F}^1(B^{3})_{i-1/2,j,k-1}+
 \bar{F}^1(B^{3})_{i-1/2,j,k}-\\
 \bar{F}^3(B^{1})_{i,j,k-1/2}- \bar{F}^3(B^{1})_{i-1,j,k-1/2}]
\end{multline}
  \begin{multline}
 (\bar{E}^3)_{i-1/2,j-1/2,k}=  0.25[\bar{F}^2(B^{1})_{i-1,j-1/2,k}+
 \bar{F}^2(B^{1})_{i,j-1/2,k}-\\
 \bar{F}^1(B^{2})_{i-1/2,j,k}-\bar{F}^1(B^{2})_{i-1/2,j-1,k}]
\end{multline}
In addition, \hammer{} upwinds these electric fields to add extra dissipation, which prevents unphysical oscillations in multi-dimensional flows (e.g. \citealt{Gardiner2005, White2016}).

To convert the face-centered magnetic field components to cell-centered magnetic field components we take volume-weighted averages in internal coordinates:

 \begin{multline}
 (\bar{B}^1)_{i,j,k}=\frac{(\bar{B}^1)_{i-1/2,j,k}+(\bar{B}^1)_{i+1/2,j,k}}{2}
\end{multline}
 \begin{multline}
 (\bar{B}^2)_{i,j,k}=\frac{(\bar{B}^2)_{i,j-1/2,k}+(\bar{B}^2)_{i,j+1/2,k}}{2}
\end{multline}
 \begin{multline}
 (\bar{B}^3)_{i,j,k}=\frac{(\bar{B}^3)_{i,j,k-1/2}+(\bar{B}^3)_{i,j,k+1/2}}{2}
\end{multline}

\subsection{Variable Inversion}
In \hammer{}, the conversion of conserved variables $\myvec U$ to primitive variables $\myvec p$ is performed using a 2-dimensional Newton-Raphson root-finding method \citep{Noble2006} or Aitken acceleration scheme \citep{Hamlin2013, Newman2014}. To provide a backup inversion method for primitive variable recovery if the energy-based inversion method(s) fail, \hammer{} also evolves the entropy equation (eq. \ref{eq:entr}), which is then used in place of the energy equation (eq. \ref{eq:ener}) to recover the primitive variables \citep{Noble2009}. To our experience, most of the inversion failures occur in highly magnetized regions such as jets where the lack of shock heating and magnetic reconnection does not have significant effects on the dynamics since $u_g$ is subdominant (e.g. \citealt{Chatterjee2019}). In fact, evolving the entropy gives us the conservative (lower limit) estimate of the internal energy. Note that since the total entropy per particle is given by $S=\frac{1}{\gamma - 1}\ln{\kappa}$, evolving $U=\kappa \rho u^t$ instead of $U=S \rho u^t$ does not conserve entropy to machine precision. However, it avoids numerical issues (e.g. \citealt{Sadowski2016}) associated with evolving logarithms, which, to our experience, makes the scheme more robust.

\subsection{Density and Internal Energy Floors}
Magnetic field lines in the polar regions (those that thread the black hole's event horizon) become devoid of matter: the gas is either expelled outwards or consumed by the black hole. Because GRMHD equations are vacuum-phobic, \hammer{} artificially injects mass and internal energy in the drift frame of the jet if the density or internal energy drop below a certain threshold \citep{Ressler2017}. This avoids both a runaway in velocity, which can occur when mass is inserted in the fluid frame, and a drag on the field lines, which can occur when mass is inserted in a zero angular momentum observer (ZAMO) frame \citep{Mckinney2012}. For this particular work we enforce $b^2/\rho<50$, $b^2/u_g<750$, $\rho>10^{-7}/r^2$, and $u_g>10^{-9}/r^{10/3}$ for the gas density $\rho$ and internal energy $u_{g}$.

\section{Numerical Optimizations}
\label{ch1:sec:Optimizations}
\hammer{} is written in C and triple parallelized: (i) CUDA or OpenCL does most of the computations on GPUs or AVX accelerated CPUs, (ii) OpenMP routines handle communication and gridding, and (iii) non-blocking MPI handles the transfer of boundary cells between nodes. NVLINK is used for transfers between GPUs on a single node and GPU-DIRECT for MPI transfers between GPUs on different nodes. We found this improves the performance by up to $20\%$ on large GPU clusters.

\hammer{} employs limited com\-mu\-ni\-ca\-tion-com\-pu\-ta\-tion overlap by (un)packing the data into send/receive buffers and applying boundary conditions in parallel with (non-blocking) MPI send/receive calls for other SMR/AMR blocks. This is made possible by using a separate CUDA stream for each block. \hammer{} also features fully non-blocking MPI-parallelized I/O, allowing substantial overlap between computation and data transfer. Data is stored in binary format. Since individual dump files can reach $10^2-10^3$ GB in size, we use OpenMP parallelized C kernels coupled to a Python script to post-process the data. These analysis tools also allow the user to load in data at reduced resolutions for post-processing.

Memory to store all variables (primitive, conserved, fluxes etc.) is dynamically allocated on both the CPU and GPU at the start of each run. During each (de-)refinement and load balancing step, C pointers are used to keep track of the physical location in memory of each SMR/AMR block. This avoids needing to allocate and deallocate large chunks of memory.

\subsection{GPU Acceleration in CUDA}
\hammer{} is the first GPU-accelerated GRMHD code that has ran on 1000s of GPUs. While CPUs spend the largest portion of their silicon on control logic and caches, GPUs spend most of their silicon on floating point power. This presents unique challenges in optimizing complex MHD codes (e.g. \citealt{Grete2019}). Hence, \hammer{} follows the philosophy of keeping the code as simple as possible, avoiding abstract concepts such as classes, extra function layers and 3-dimensional arrays, which makes it easy for the CUDA compiler to generate highly efficient code. In comparison to the GPU-accelerated (GR)MHD codes KHARMA (Prather et al, in prep) and K-ATHENA \citep{Grete2019}, which use KOKKOS \citep{Kokkos2014}, we chose to implement the GRMHD equations directly into NVIDIA's CUDA. KOKKOS is a performance portability toolkit that can compile existing C code on various architectures. While KOKKOS increases the performance portability of a code across various architectures, it is unlikely to be able to match the performance of a highly optimized CUDA code that makes extensive use of NVIDIA specific optimizations.

It was challenging to develop the GPU version of \hammer{} due to the large code size (over 40,000 lines of C code and 10,000 lines of CUDA kernel code). This made it difficult to fit all necessary instructions and variables into the GPU's register space. Failing to fit the code into the available register space would slow down the performance by an order of magnitude, because the GPU would need to use the much slower GPU RAM to read and store temporary data. Kernels in many conventional codes are rather small and do not fill up the GPU register space easily. GRMHD codes, however, need to store metric components of curved space-time of a black hole and require non-linear root finding to convert conserved to primitive quantities \citep[e.g.][]{Noble2006}. For this reason, fitting the GRMHD equations required many profiler-informed optimizations based on an iterative trial-and-error approach. It was, for example, beneficial to split the code into extra kernels to recalculate certain quantities instead of storing them in temporary variables using register space. This profiler-based optimization strategy has substantially increased the performance of \hammer{} since 2016. The first 2016 version of \hammer{} was memory bandwidth limited because of significant register spillover to global memory. Eventually we reduced the required memory bandwidth and increased \hammer{}'s performance 5-fold by rewriting various functions to fit into the available register space. We do not make extensive use of shared memory (e.g. \citealt{Begue2022}) as we found that this slows down computation compared to a more efficient use of register space. Shared memory is only used in \hammer{} to calculate the minimum timestep in each mesh-block through parallel reduction.

There exists a trade-off, different for each GPU generation, between making GPU kernels too big and complex to fit into the available register space and having too many small kernels with the added overhead. On the previous generation of GPUs (NVIDIA Kepler architecture) it was, for example, beneficial to split the code into extra kernels to recalculate certain quantities instead of storing them in temporary variables using register space.  On the present generation of GPUs (NVIDIA Pascal/Volta/Ampere architectures) it is beneficial to make some kernels a bit bigger, due to the larger register space. We achieve best performance when we split a timestep into $\sim 7$ separate kernels.

\begin{figure}
\begin{center}
\includegraphics[width=\columnwidth,trim=0mm 0mm 0mm 0,clip]{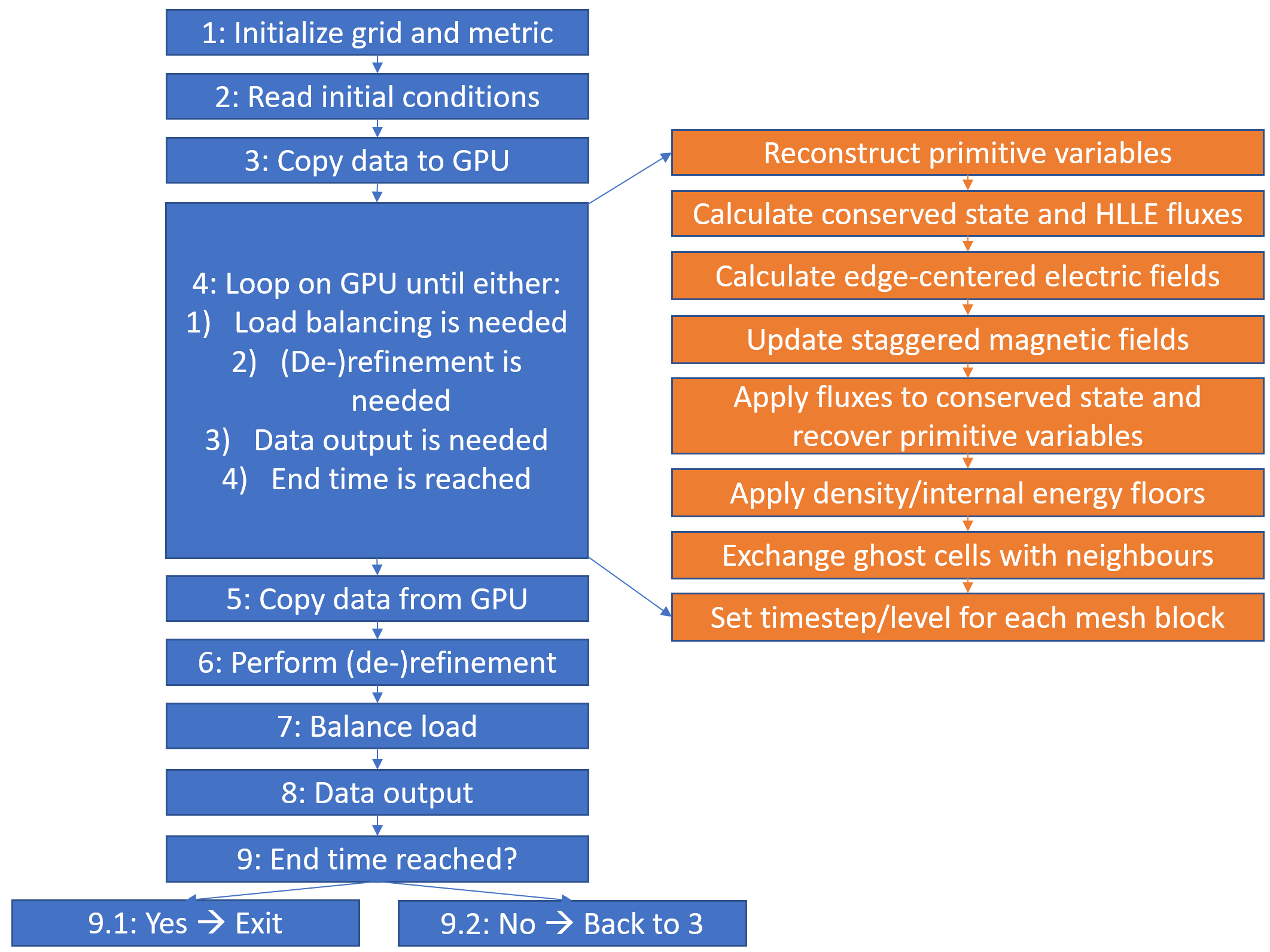}
\caption{A code diagram of \hammer{}. On the left (blue) are all tasks that are peformed using OpenMP parallelized C functions. On the right (orange) are all tasks that are performed using GPU-accelerated CUDA kernels. Practically all of the computation is GPU-accelerated and the non-accelerated parts in \hammer{} typically do not form a performance bottleneck.
}
\label{fig:code}
\end{center}
\end{figure}

All functions necessary to evolve the state of a cell from $t^{n}$ to $t^{n+1}$ are GPU accelerated. These include reconstruction routines, calculation of HLL fluxes, conserved-to-primitive variable inversion, setting of density/internal energy floors, and exchange of boundary/ghost cells. The CPU handles all the necessary logic to manage the transfer of boundary cells, fluxes and electric fields between the neighbouring blocks and performs the (de-)refinement and load balancing steps including the divergence-free prolongation and restriction of magnetic fields \citep{Balsara2001}. We have summarized the parallelization strategy of \hammer{} in a code diagram (Fig.~\ref{fig:code}).

\subsection{AVX Acceleration in OpenCL}
In addition to the `normal' OpenMP+MPI C based CPU code and GPU accelerated CUDA code, \hammer{} maintains a legacy OpenCL branch that can make use of AVX vectorization on CPUs. This effectively leverages the 512 bit wide AVX vector registers in the latest generation of Intel Skylake CPUs. Without any CPU-specific optimizations to our GPU kernels (e.g. with the CUDA kernels only ported to OpenCL) \hammer{} achieves $\sim 12\%$ saturation of the CPU's (16 Core AVX-512 2.8 GHz Skylake) theoretical FP64 throughput and gain a factor $\sim 3$ speed-up compared to standard OpenMP code.

\subsection{Adaptive Mesh Refinement}
\label{sec:amr}
Adaptive mesh refinement (AMR, see \citealt{Berger1989, Balsara2001}) allows \hammer{} to focus the resolution on regions of interest. While not bringing significant advantages for simulating thick accretion disks, which span most of the computational domain, AMR can speedup computations requiring high-resolution of small-scale features that fill a small fraction of the computational volume. For instance, for problems such as thin accretion disks, tidal disruption events, and large-scale collimated jets, AMR can reduce the number of required cells by orders of magnitude.

AMR can also speed up the calculations in another way: by \emph{reducing} the spatial resolution in the blocks closest to the black hole and thereby increasing the timestep. Such a reduction in resolution is feasible, because GRMHD modeling has consistently shown that logarithmically-spaced spherical grids typically sufficiently resolve the turbulence closest to the black hole (e.g., within two event horizon radii), while resolving the outer regions in the accretion disk and relativistic jets remains a challenge (e.g.\ \citealt{Mckinney2006, Liska2018C, Porth2019}).

We designed the AMR framework in \hammer{} from the ground up for performance and scalability. We make use of an oct-tree based approach, where any parent block can be split in either 2, 4, 6 or 8 child blocks (see Sec.~\ref{ch1:sec:Pole}). Compared to a patch-based AMR approach, we do not need to evolve the underlying parent blocks for each refined layer and, instead, can directly transfer boundary cells from neighbouring coarse to fine layers of the grid (and vice-versa). This drastically reduces the required inter-node MPI bandwidth, which presents the main performance bottleneck for our simulations. Since the relative inter-node MPI bandwidth is expected to decrease for the next generation of GPU clusters, oct-tree based AMR is an attractive approach for the next generation(s) of supercomputers.

Since every problem is different, \hammer{} allows the user to implement an arbitrary refinement criterion. For example, \citet{Liska2018C, Liska2019b} used a cutoff on the maximum density in each block as the refinement criterion. Blocks satisfying the chosen refinement criterion will be refined to the highest AMR level allowed for a problem. Neighboring blocks (adjacent along the faces and edges) for each refined block, which do not satisfy the refinement criterion, are refined to 1 AMR level below that of the refined block. These nesting conditions prevent resolution jumps by factors greater than $2$ along block faces and edges.

We use prolongation and restriction operators similar to those described in \citet{BergerOliger1984, Berger1989} for cell-centered variables and \citet{Balsara2001} for face-centered variables. These operators upscale/downscale the resolution of the grid during (de-)refinement steps, and reconstruct primitive variables in ghost cells at coarse-fine boundaries. During prolongation we use MINMOD limited gradients calculated on the coarse mesh to upscale the resolution of the variables to a finer mesh. By using a more diffusive MINMOD limiter we prevent numerical issues at refinement boundaries associated with applying a 1D reconstruction scheme to a 3D stencil (see also \citealt{Stone2020}). During restriction, we take the volume or area weighted averages on the fine mesh to reduce the resolution of the variables on the coarse mesh. Prolongation and restriction can be performed on either the primitive variables or conserved quantities. The former is faster and more robust (e.g. does not lead to states with $v>c)$, while the latter is energy and angular momentum conserving up to machine precision. In general, if the refinement criterion is set such that no (de-)refinement happens in the area of interest we recommend to perform prolongation and restriction directly on the primitive variables. 

In addition, as described in \citet{Berger1989}, we replace at the end of a timestep the fluxes on the coarser mesh with the area-averaged fine mesh fluxes and edge-averaged electric fields. This is necessary since the sum of the fluxes on the fine mesh might not match to machine precision the flux on the coarse mesh. This refluxing guarantees conservation of mass, energy, and angular momentum, in addition to the divergence free evolution of magnetic fields. While refluxing of cell-centered quantities only involves 6 faces, refluxing of the edge-centered electric fields involves, in addition to the 6 faces, 12 cell edges. In general, we pick the cell edge at the highest refinement and timestepping level as being `dominant'. The electric fields on neighbouring edges are subsequently synchronized with the `dominant' edge. To make sure that this constrained-transport algorithm is implemented correctly, we periodically verify consistency up to machine precision between the staggered magnetic field components at cell boundaries of neighbouring meshblocks.

\subsection{Handling of the Polar Region}
\label{ch1:sec:Pole}

Spherical grids are perhaps the most natural and efficient grids to study accretion disks. They naturally follow the geometry of the disk rotating around the black hole. The spherical shape of the event horizon in the Kerr-Schild foliation naturally matches that of the grid's inner boundary, $r = R_{\rm in}$. In addition, spherical grids support a logarithmic spacing in the radial coordinate, naturally providing high resolutions close to the black hole where the timescales are short, and lowering the resolution progressively as one moves out and the timescales become longer. This is advantageous compared to, for example, Cartesian grids, as one does not need to use many AMR layers to get the typically required 2 to 5 orders of magnitude scale separation between inner and outer boundary of the grid. This in turn allows for the use of relatively large AMR block sizes, minimizing the number of boundary cell transfers and leading to excellent parallel scaling (Sec.~\ref{ch1:sec:Scaling}).

A drawback of spherical grids is the polar coordinate singularity, which requires special treatment. We have implemented transmissive boundary conditions across the singularity to minimize dissipation and verified that the grid is robust for the study of tilted accretion disks and jets (see Appendix of \citealt{Liska2018A} for details). Even with these improvements, spherical grids still suffer from cell `squeezing' near the pole in the azimuthal direction. This causes the \citet{Courant1928} condition (which limits the timestep to the wave-crossing time of single cell) to limit the global timestep more than for a Cartesian with an equivalent resolution. Since close to the polar singularity the cell aspect ratio $r\Delta \theta/(r\sin\theta \Delta\varphi) \simeq \Delta\varphi^{-1} = N_\varphi/(2\pi)$ increases proportional to the azimuthal resolution, the timestep for a spherical grid will, empirically, be a factor $ \sim N_{\varphi}/(2\pi)$ times smaller than that for a Cartesian grid with the same effective resolution.\footnote{More precisely, the reduction factor will be $3$ times smaller, since the time step in \hammer{} is set as the harmonic mean of the time steps in each of the three dimensions, $1/\Delta t = 1/\Delta t_1+1/\Delta t_2+1/\Delta t_3$. Hence, for cubical cells, $\Delta t \approx (1/3)\Delta t^1 \approx (1/3)\Delta t^2 \approx (1/3)\Delta t^3$.} As a result, in 3D the cost of simulations on a spherical grid increases as $\propto N^5$ instead of the typical $\propto N^4$ for Cartesian grids, making high resolution (GR)MHD simulations on spherical grids particularly expensive. 

\begin{figure}
\begin{center}
\includegraphics[width=\columnwidth,trim=0mm 0mm 0mm 0,clip]{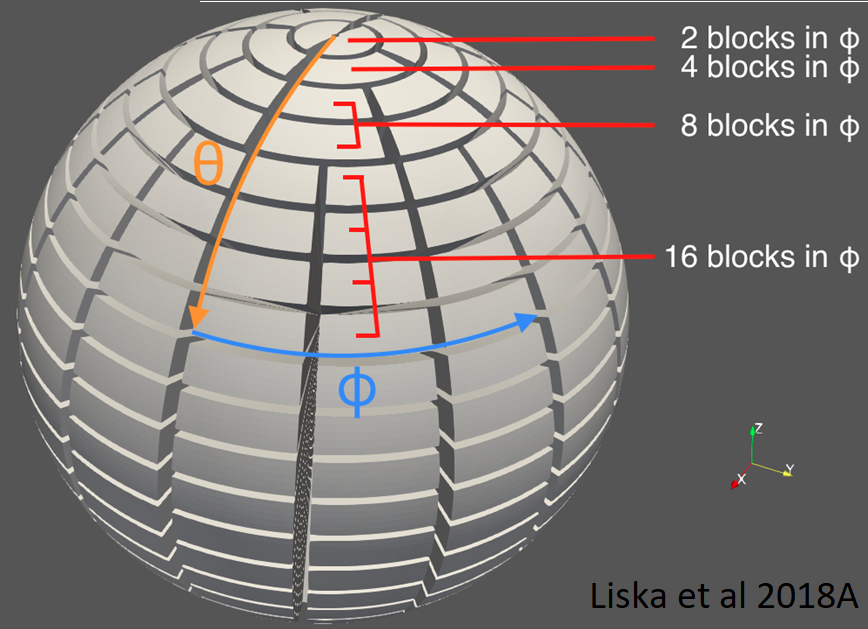}
\caption{As previously presented in \citet{Liska2018A}, the spherical `base' grid in \hammer{} has a varying number of blocks in $\varphi$. This prevents cell squeezing near the pole, speeding up high-resolution GRMHD simulations by up to $2$ orders of magnitude without any loss in accuracy. AMR can increase the resolution of all blocks in this grid, including those adjacent to the polar axis.
}
\label{ch1:fig:grid}
\end{center}
\end{figure}

There are three popular approaches to avoiding the squeezing of the cells in the $\varphi$-direction. One is the use of Cartesian grids (e.g \citealt{Porth2019}). The advantage of this method is that it is symmetry-agnostic. Disadvantages include the added computational cost because the grid does not conform to the shape of the black hole or the orbital motion of the gas. Such grids can also require many AMR levels \citep[e.g., $8$ AMR levels,][]{2019arXiv190610065D}, which can reduce parallel scaling efficiency, a point of concern especially for GPU-based systems with a low ratio of MPI interconnect bandwidth to computation power. The second approach (e.g \citealt{Stone2020}) is to use static mesh refinement (SMR) to derefine the grid near the pole in all 3 dimensions, ($r$, $\theta$, $\varphi$). The lower polar resolution reduces the computational cost and is excellent for applications focused on the physics of the equatorial accretion flow. However, the larger cell size in the $r$- and $\theta$-directions also makes it more difficult to resolve polar structures such as jets or tilted disks. Currently, most GRMHD codes use a third strategy which involves enlarging the innermost few cells in the $\theta-$direction \citep{Tchekhovskoy2011}. While this approach works well at low resolutions, it becomes ineffective at higher resolutions since the relative size of the innermost cells increases proportional to the $\phi-$resolution. For example, at the grid resolution of $128^3$ cells the innermost cell needs to be enlarged by a factor $\sim 10$, while at the resolution of $1024^3$ cells it would need to be enlarged by a factor $\sim 100$. This makes grid cylindrification unsuitable for high resolutions.

\subsubsection{External SMR}
\label{sec:external-smr}
In \hammer{}, we attempted to combine the best aspects of both of the above approaches. We use static mesh refinement (SMR) to avoid the squeezing of cells in the $\varphi$-direction \citep{Liska2018A}. For this, \hammer{} derefines the base layer of the grid in the $\varphi$-direction within $30^{\circ}$ of the pole. This maintains the full resolution in the $r$- and $\theta$-directions, leading to a uniform cell aspect ratio (within a factor of $2$) across radial shells (Fig.~\ref{ch1:fig:grid}). As an example, our fiducial simulation (which uses base grid with, effectively, $1728 \times 576 \times 1024$ cells; Sec.~\ref{ch1:sec:Scaling}) has a $\varphi$-resolution of 256 cells for $0^{\circ}<\theta<15^{\circ}$, 512 cells for $15^{\circ}<\theta<30^{\circ}$ and the full 1024 cells for $30^{\circ}<\theta<90^{\circ}$. We refer to this approach as ``external SMR'', to distinguish it from ``internal SMR'' discussed below (see Sec.~\ref{sec:internal-smr}). 

A similar approach to reduce the resolution near the pole is utilized in the radiation hydrodynamics FORNAX code \citep{Skinner2019}. However, to our knowledge, \hammer{} is unique among (GR)MHD codes in its ability to combine 1-dimensional SMR with 3-dimensional AMR. When refined with AMR, all parent blocks in our SMR-enhanced base grid are split into 8 child blocks, except the blocks touching the polar axis. These blocks are split into 6 child blocks, where the 2 child blocks along the pole maintain their parent's $\varphi-$resolution and the 4 child blocks further away from the pole are refined in all 3 dimensions. This way the cells near the pole are kept approximately cubic by the (effective) addition of a derefinement layer in the $\varphi-$ direction for each additional AMR layer. 

\subsubsection{Internal SMR}
\label{sec:internal-smr}
Typically, \hammer{} uses external SMR to reduce the $\varphi$-resolution down to $N_{\varphi} \sim 64-256$ cells near the pole. Bringing down the $\varphi$-resolution to $N_{\varphi} \sim 16$ cells would require $2-4$ extra refinement layers featuring exceedingly small block sizes. This would oversaturate the available memory and MPI interconnect bandwidth, leading to a serious performance bottleneck on GPU clusters. To remedy this limitation, \hammer{} also implements $\varphi$-derefinement internally on the block level, which we refer to as internal SMR. Depending on the distance of a cell from the pole, internal SMR spatially averages the fluxes and conserved quantities of the cell over $2^n$ cells in the $\varphi$-direction within $2^{n_{\rm max}-n}$ cells from the pole such that the cell aspect ratio remains close to unity. Here, $n$ is the internal SMR level and $n_{\rm max}$ is the number of internal SMR levels. This can reduce the $\varphi$-resolution by an additional factor $4-16$, which is sufficient to completely eliminate cell `squeezing' near the pole while exploiting the computational advantages of larger-sized blocks. To maintain higher order spatial accuracy, the spatial reconstruction method is modified accordingly to compensate for the (effectively) increased cell spacing, making internal SMR equivalent to external SMR.

\subsection{Local Adaptive Timestepping}
\label{ch1:sec:LAT}
\hammer{} features a local adaptive timestep (LAT). The typical approach for this in AMR codes is to use smaller timesteps for higher-level AMR layers (aka hierarchical timestepping). In addition, \hammer{} exploits the fact that on a logarithmic grid, cell size increases with radius, even without any spatial refinement layers. Based on the local \citet{Courant1928} condition, LAT allows \hammer{} to determine the time step size (in steps of a factor of 2) for each block (this includes the block's ghost cells) independently from the block's spatial refinement level. To guarantee $2^{nd}$ order convergence in time, we typically evolve the coarser grid first such that we can estimate (by linear interpolation in time) the ghost cell variables on the fine grid. Subsequently, in a flux correction step (see also section \ref{sec:amr}), the flux difference between the fine and course grid is applied to the coarse grid.

LAT leads to a factor of $3{-}10$ reduction in the number of timesteps, depending on the structure of the adaptive grid. This manifests itself as a factor $2-3$ speedup due to extra overhead associated with parallel scaling (see Sec.~\ref{ch1:sec:Balance}). In general, grids with many refinement layers further away from the black hole achieve the most benefit from LAT, while grids with no AMR only manage to reduce the number of timesteps by a factor $\sim$few. In contrast, using hierarchical timestepping, which couples the timestep to the spatial refinement level, would lead to a smaller speedup of $10-30\%$ since for most problems most blocks reside at the highest spatial refinement level. In addition to speeding up computations, LAT also increases the numerical accuracy by reducing the number of conserved-to-primitive variable inversions in the outer grid and thereby reducing the noise generated by the inversions (see the appendix in \citealt{Chatterjee2019}). This is especially pronounced in large-scale jets, where the calculations due to the large difference between the magnetic energy density and rest mass energy density are prone to inversion errors.

\subsection{Load Balancing}
\label{ch1:sec:Balance}
We use a z-order space filling curve for load balancing in order to keep neighbouring blocks in the grid physically close to each other on a cluster (e.g., on the same or neighbouring nodes). Depending on the problem specifics (such as the number of spatial and temporal refinement levels) and the architecture of the cluster (fat-tree or 3D-torus), we found that changing the fastest moving index in the space filling curve or switching to a row-major order at the $0^{\rm th}$-level may significantly improve performance. AMR (de)refinement is performed every $10^3-10^4$ timesteps. This is similar to one orbital period at the black hole event horizon, hence sufficient to capture the dynamical evolution of the accretion disk and jet. 

We note that LAT brings about two technical difficulties. First, it significantly complicates load balancing. If one were to keep the number of timesteps per GPU equal and naively follow the space-filling curve, a huge imbalance in memory consumption could occur, causing some GPUs to run out of memory. Second, it makes the synchronization of fluxes between different time levels more difficult: at every timestep the conserved fluxes and electric fields need to be synchronized between the fine and coarse time levels on block faces and edges to guarantee energy/momentum conservation and the divergence-free evolution of magnetic fields \citep{Balsara2001}. For this, by the end of each full timestep the flux difference between the fine and coarse layers is added as a correction to the coarse block's boundary cells and thus an additional conserved to primitive variable inversion needs to be performed for these boundary cells. This tightly couples the fine and coarse layers, potentially leading to an unbalanced load that slows down computation. We now discuss how we alleviate both of the issues.

\begin{figure*}
\begin{center}
\includegraphics[width=1.0\linewidth,trim=0mm 0mm 0mm 0,clip]{AMR_v2.jpg}
\end{center}
\caption{GRMHD simulation of a thin accretion disk of aspect ratio $h/r=0.02$ initially threaded by a purely toroidal magnetic field and tilted by $65^{\circ}$ relative to the (horizontal) equatorial plane of the black hole. The frame-dragging by the spinning black hole rips the accretion disk apart into inner and outer subdisks as can be seen in the rendering of density iso-contours in the left panel. The black hole spin points along the positive z-direction. The right panel shows a vertical slice through the density. The top left zoom-in inset shows that the inner regions of the accretion flow reorient themselves parallel to the BH midplane as it undergoes Bardeen-Petterson \citet{bp75} alignment, the first demonstration of this effect in GRMHD in the absence of large-scale poloidal magnetic field (see also \citealt{Liska2018C, Liska2019b}), solving a 40-year-old problem \citep{bp75}. The bottom right inset illustrates that the small-scale turbulent structure of the accretion flow is very well resolved.
An animation of this figure is available online. A higher resolution version is also available as a \href{https://www.youtube.com/watch?v=rIOjKUfzcvI}{YouTube movie}. The animation runs from $t \sim 10,000 r_g/c$ to $t \sim 138,000 r_g/c$ and has a real-time duration of 104 seconds.} 
\label{ch1:fig:AMR}
\end{figure*}

To prevent an imbalance in memory consumption between GPUs, \hammer{} uses a load balancing strategy that attempts to keep both the number of timesteps and the number of blocks at each timelevel per GPU constant, while adhering as closely as possible to the utilized space filling curve. This can be achieved by loading in on every GPU a similar number of computationally intensive (with many timesteps) and non-intensive (with few timesteps) blocks by load balancing the grid separately for each timelevel, and, using the next timelevel to `fill' up the imbalance created during load balancing of the previous timelevel. To prevent GPUs from running out of memory during e.g. (de)refinement steps (each 16 GB GPU can only fit a grid of approximately $\sim 256^3$ cells), \hammer{} may in rare cases also perform intermediate load balancing steps or may reduce the number of timelevels (if it determines that this indeed reduces the inter-node memory imbalance).

\section{Performance Measurements}
\label{ch1:sec:Scaling}
In this section, we describe the physical and numerical setup of the GRMHD simulation we used as a benchmark. This simulation is illustrated using a volume rendering and a transverse slice through density in Figure \ref{ch1:fig:AMR}. It ran in 2019 on then the world's most powerful supercomputer, OLCF Summit, which is powered primarily by $27648$ NVIDIA V100 GPUs, each with 16 GB of HBM2 RAM. We also discuss how we measure the performance and determine the scaling efficiency of \hammer{}. 

\subsection{Physical Setup}
We consider a thin accretion disk with scale height $h/r=0.02$ around a spinning black hole of spin $a=0.94$ tilted by $65^{\circ}$. Our previous work initialized the accretion disk with a poloidal magnetic flux loop \citep{Liska2018C, Liska2019b}. In contrast, here the disk is threaded by a toroidal magnetic field with an approximately uniform plasma $\beta \sim 10$ \citep{Musoke2022}. To our knowledge, this is the first GRMHD simulation of a thin disk in the absence of initial poloidal magnetic flux, making it extra challenging since a toroidal magnetic field needs a much higher resolution to be properly resolved \citep{Liska2018B}. Such a magnetic field topology is thought to be responsible for most of the luminous black hole systems in our universe, since most of such systems lack a jet, while a poloidal magnetic field guarantees the production of a jet. We discuss the physical implications of the simulation in Sec.~\ref{sec:scientific-results}.

\subsection{Numerical Grid}
\label{sec:numerical-grid}
The numerical grid stretches from the inner boundary just inside the event horizon, at $\sim r_g$, to the outer boundary at $\sim 10^5 r_g$. It has a total effective resolution of $13440 \times 4608 \times 8096$ in the disk beyond $\sim 10 r_g$. The resolution gradually drops to the base-grid resolution of $1728 \times 576 \times 1024$ at $r \sim 4 r_g$ (this increases the time step by about an order of magnitude; Sec.~\ref{sec:amr}). The cell sizes at $r=(1.5,8,20,50)r_g$ are $|\Delta r| \sim |\Delta \theta| \sim |\Delta \phi|\sim (0.0082, 0.011, 0.011, 0.034)$. The grid has $0.9-1.4 \times 10^5$ blocks of size $48 \times 48 \times 64$ cells and a total of $\sim 12-22$ billion cells. The number of blocks and cells varies throughout the evolution as AMR creates and destroys blocks to focus the resolution on the disk body as it moves through the grid. For this problem, we use 4 levels of AMR (Sec.~\ref{sec:amr}), 2 external (Sec.~\ref{sec:external-smr}) and 4 internal (Sec.~\ref{sec:internal-smr}) layers of SMR, and 5 levels of LAT (Sec.~\ref{ch1:sec:LAT}). 

Total resolution-wise, this advanced grid improves by more than an order of magnitude on our previous simulations of $0.25-1.5$ billion cells carried out on the NCSA Blue Waters supercomputer (e.g. \citealt{Liska2019b}). Consequently, this improves the resolution of the disk to $25{-}30$ cells, up from $7{-}14$ cells per scaleheight. Typically around 4-5 scale heights of the disk reside in the highest refinement level, ensuring uniform resolution of the magnetized turbulence not only near the midplane of the disk, but also further away from the midplane.

\subsection{Performance Measurements}
\label{sec:weak-scaling-tests}
The speed of GRMHD codes is usually expressed in zone-cycles/s. Each zone-cycle represents an operation that advances a single cell in the grid for one full timestep. The computational cost in terms of memory bandwidth and number of double precision floating pointing operations is de-facto constant for each zone-cycle and was measured on a single GPU with NVPROF. Thus it is trivial to calculate those metrics for our fiducial run on OLCF Summit based on the achieved zone-cycles/s. Similarly, the total memory consumption can be calculated by measuring the size of a single block and multiplying this number by the total number of blocks in the grid. In addition, we calculate the MPI bandwidth by counting the number cell transfers between AMR/SMR blocks on different nodes.

The grid described in Sec.~\ref{sec:numerical-grid} is used for the largest of our weak scaling tests, which uses $5,\!400$ V100 GPUs and requires over 25TB of GPU RAM. We use fewer blocks of the same size ($48\times48\times64$) for the smaller scaling tests. In other words, we keep the number of blocks/GPU constant by preventing \hammer{} from refining beyond a certain number of blocks (this can potentially reduce the number of AMR levels). Note that since the grid for the smaller weak scaling tests becomes smaller and may have fewer AMR layers the problem is not exactly the same and thus these benchmarks merely serve as an indication of what performance might look like for a hypothetical simulation using a smaller grid with similar characteristics. The plotted efficiency is accurate though for our fiducial simulation and represents the average performance of our code, which includes the entire application performance including I/O. 

We express in Sec.~\ref{ch1:sec:Scaling} our numerical efficiency as the achieved zone-cycles/s/GPU relative to the maximum value devised from single GPU benchmarks with a $150^3$ block size (Sec.~\ref{ch1:sec:Benchmarks}). Similarly we express the weak scaling efficiency as the achieved zone-cycles/s/node compared to this value achieved for a single node (6 V100 GPUs) using the respective grid ($\sim 20$ blocks of size $48\times48\times64$ per GPU).

\begin{figure}
\begin{center}
\includegraphics[width=\linewidth,trim=0mm 0mm 0mm 0,clip]{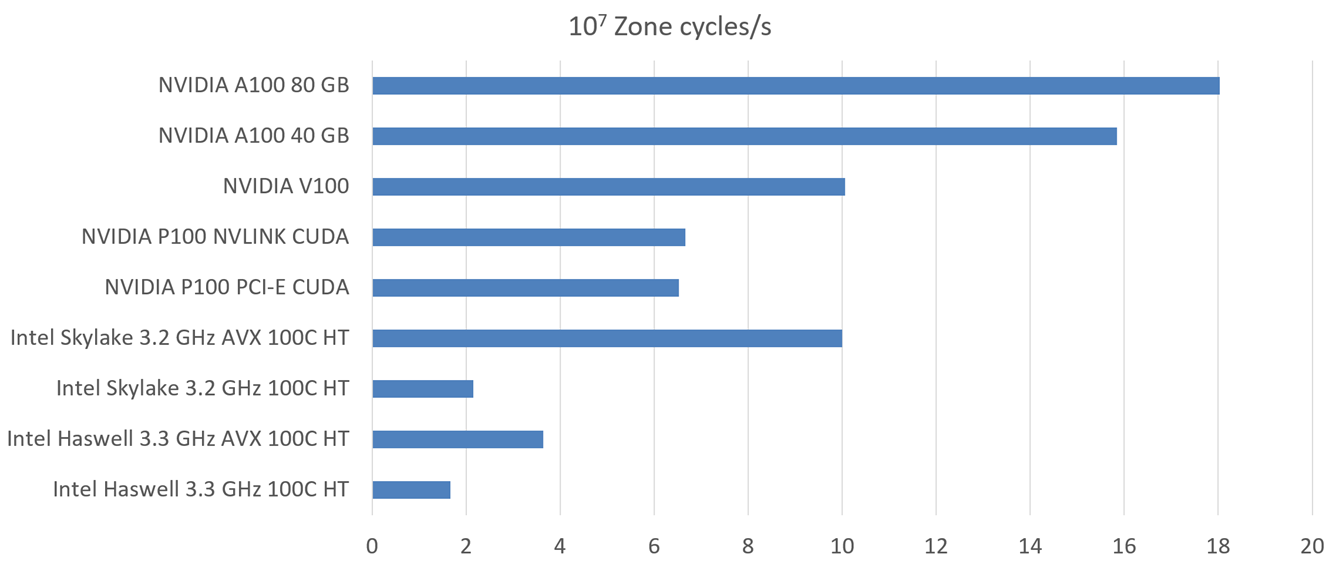}
\end{center}
\caption{The computational performance of \hammer{} in zone cycles/s on a $150^3$ uniform grid for various generations of Intel CPUs (Haswell and Skylake) and NVIDIA GPUs (Pascal, Volta and Ampere). HT indicates we make use of HyperThreading  on Intel CPUs. A zone cycle includes both the predictor and corrector step in our $2^{\rm nd}$ order accurate timestepping routine.
}
\label{ch1:fig:benchmarks}
\end{figure}

\subsection{Limitations on Scaling}
As long as the grid has a resolution of $\gtrsim 125^3$ per GPU we find that the numerical efficiency remains excellent for a wide range of problems. Lowering the resolution to $50^3$/GPU without LAT or to $\sim 100^3$/GPU with LAT gives us only $\sim 40\%$ of the GPU's maximum performance in zone-cycles/s. The per-GPU resolution limitation arises because a GPU needs to keep $\sim 10^5$ CUDA threads occupied, which is not possible for smaller grids, especially when LAT reduces the total number of timesteps by a factor of $\sim 10$. To allow for efficient load balancing this grid needs to be divided such that there at least $\gtrsim 15$ blocks per GPU.  Using $\lesssim 7$ blocks per GPU leads to a factor $2-3$ performance decrease. Another constraint comes from the block size, which needs to exceed $\gtrsim 50^3$ cells, such that inter-node MPI transfers do not become prohibiting. For block sizes of $\sim 50^3$ cells we find that on average $\sim 25 \%$ of the MPI interconnect bandwidth is saturated for our fiducial run (Table \ref{efficiency_table}). However, since some nodes can be more MPI-intensive than others, this number can be much higher for certain nodes and makes the MPI bandwidth the main performance bottleneck for our fiducial run. Reducing the block size to $\sim 26^3$ while keeping the total number of cells constant decreases the performance by a factor $2-3$. Furthermore, the available HBM2 RAM on each $16$GB GPU can only fit a $256^3$ grid.

These constraints make it imperative that every grid has to be fine-tuned to the respective system . In our case the simulation has run $\sim 10\%$ of the time on $3600$ V100 GPUs, which is the minimum amount of GPUs we need due to RAM constraints, and $\sim 90\%$ of the time on $5,\!400$ V100 GPUs, above which there are not enough blocks to effectively load balance the grid. We find that within this range the number of zone-cycles-per-second per GPU remains constant to within $10\%$. 

\begin{figure}
\begin{center}
\includegraphics[width=\columnwidth,trim=0mm 0mm 0mm 0,clip]{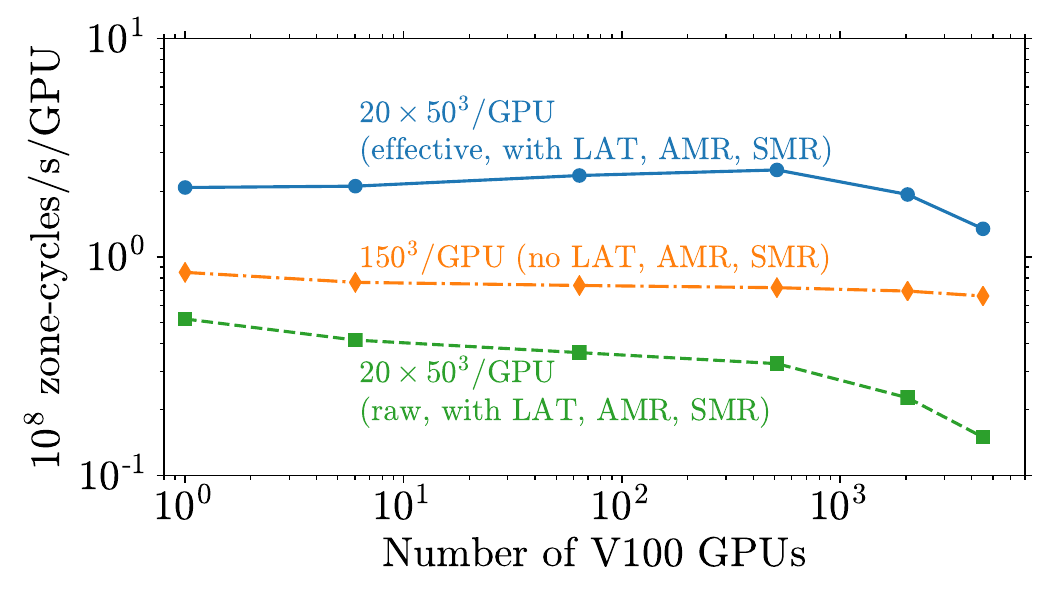}
\end{center}
\caption{\hammer{} shows excellent weak scaling for both simple and complex grids. For simple grids, it shows $\sim80\%$ numerical efficiency on 900 OLCF Summit nodes = 5,400 GPUs for a single block of $150^3$ cells per GPU, without using AMR, SMR, or LAT (dashed-dotted orange line). For more complex grids, which use AMR and SMR, with 20 blocks of size $48\times48\times64$ $\sim 50^3$ cells per GPU, the efficiency at 900 nodes drops to $\sim60\%$ (not shown for brevity). While using local adaptive timestepping (LAT) decreases the raw parallel efficiency to $\sim35\%$ (dashed green line), it also effectively speeds up the simulations by a factor of $4$ (on 1 GPU) to $9$ (on 5,400 GPUs, solid blue line), leading to an effective numerical efficiency of nearly $\sim 200\%$ on 5,400 GPUs. }
\label{ch1:fig:scaling}
\end{figure}

\section{Performance Results}
\label{sec:performance-results}
In this section we demonstrate excellent single GPU performance, time-to-solution, scalability and peak performance for the thin accretion disk problem presented in Sec.~\ref{ch1:sec:Scaling}.

\subsection{Single CPU/GPU Benchmarks}
\label{ch1:sec:Benchmarks}
Figure~\ref{ch1:fig:benchmarks} compares the computational performance in zone-cycles/s of \hammer{} on various GPU and CPU architectures for a simple test configuration without using AMR, SMR or LAT and block size of $150^3$. On a single NVIDIA V100 GPU \hammer{} attains $\sim 1,0\times10^8$ zone cycles/s, which corresponds to a factor $\sim 5$ speedup compared to a 28 core Intel W3175X Skylake CPU (clocked at 2.8 GHz with 16 cores active), which attains $10 \times 10^5$ zone cycles/s/core using AVX-512 vectorization. Note that a zone cycle includes the predictor and corrector steps in our $2^{\rm nd}$ accurate timestepping routine. We profiled our code using NVPROF on a single GPU and manage to saturate $25\%$ of FP64 throughput and $57\%$ of memory bandwidth without any register spillover on a single NVIDIA V100 GPU for a $150^3$ block. We consider this a remarkable result, especially since \hammer{} contains more than 10,000 lines of CUDA kernel code.

\begin{table}
\begin{center}
\caption{Time-averaged absolute and relative performance of \hammer: floating point power (FP64), GPU DRAM bandwidth, GPU DRAM consumption and MPI bandwidth for our fiducial run on $5,\!400$ V100 GPUs of the OLCF Summit supercomputer. This demonstrates that \hammer{} does not have a `single' bottleneck resource, making efficient use of OLCF Summit. \label{efficiency_table}}
\begin{tabular}{|c|c|c|}
\hline
&  \textbf{Abs. perf.} & \textbf{Rel. perf.}\\
\hline
FP64 rate & $2.2$ PFLOP/s & $5.3\%$ \\
\hline
GPU DRAM bandwidth & $588$ TB/s & $12.1\%$\\
\hline
GPU DRAM consumption & $25$ TB & $28\%$ \\
\hline
MPI bandwidth & $5.5$ TB/s  & $24\%$\\
\hline
\end{tabular}
\end{center}
\end{table}

\subsection{Peak Performance}
From the total number of zone-cycles/s achieved in our fiducial run we can easily calculate the peak performance on $5,\!400$ V100 GPUs as described in Sec.~\ref{sec:weak-scaling-tests}. The results are summarized in Table~\ref{efficiency_table}, which shows that \hammer{} uses a significant fraction of both the FP64 power, GPU memory bandwidth, GPU memory size and MPI bandwidth. This leads us to conclude that \hammer{} is very well optimized across the board and makes efficient use of the resources available on OLCF Summit. Furthermore, it shows that while the MPI and GPU memory bandwidths are the current bottlenecks, implementation of new floating point heavy physics such as radiation into \hammer{} may cause the floating point capacity to become the bottleneck resource in the future.

\begin{table*}
\begin{center}
\caption{Performance metrics of \hammer{} in 4 astrophysical scenarios. TD-4608 is the fiducial run of a tilted thin accretion disk described in this article, which makes use of 3 levels of AMR. TD-1728 is an aligned thin accretion disk that makes use of 3 levels of AMR \citep{Liska2022}. EHT-192 is a geometrically thick accretion disk that does not use AMR \citep{Porth2019}. It represents a typical medium-resolution GRMHD simulation part of the Event Horizon Telescope Collaboration (EHTC) GRMHD simulation library. EHT-2304 is a similar simulation of a geometrically thick accretion, but performed at an extremely high resolution without AMR \citep{Ripperda2022}. The effective number of zone-cycles and zone-cycles/s assumes that all cells are evolved at the highest LAT level. Thus the ratio between the number of effective and real zone-cycles equals the factor by which LAT reduces the number of timesteps. The $LAT \times GPU$ speedup gives the factor by which LAT and GPUs speed up \hammer{}. We compare against a 20 core Skylake CPU, on which the non-AVX-vectorized version of \hammer{} has a performance of $\sim 3.5 \times 10^6$ zone-cycles/s. SMR/AMR speedup ($ \# Timesteps$) gives the cost reduction factor in the number of timesteps compared to a spherical grid without SMR that uses cylindrification \citep{Tchekhovskoy2011} around the polar axis such that it is not limited by the Courant condition in the $\varphi-$direction up to a resolution of $N_{\varphi}=128$. For $N_{\varphi}>128$ we assume that $\Delta t^3=\frac{128}{N_{\varphi}}\Delta t^2$ and calculate the timestep as $\Delta t \propto 1/(1/\Delta t^1+1/\Delta t^2+1/\Delta t^3)$ (Sec.~\ref{ch1:sec:numerical_methods}). AMR speedup ($ \# Cells$) similarly gives the reduction factor in the number of cells, which is defined as the ratio between the true number of cells in the grid and the number of cells in a unigrid at the effective resolution. Multiplying all of these factors gives the total speedup.
\label{performance_table}}
\begin{tabular}{|c|c|c|c|c|}

\hline
&  \textbf{TD-4608} & \textbf{TD-1728} & \textbf{EHT-192} & \textbf{EHT-2304}\\
\hline
Base resolution & $1680 \times 576 \times 1024$ & $1020 \times 432 \times 288$ & $304 \times 192 \times 192$ & $5376 \times 2304 \times 2304$\\
\hline
Effective resolution & $13440 \times 4608 \times 8192$ & $4080 \times 1728 \times 1152$ & $304 \times 192 \times 192$ & $5376 \times 2304 \times 2304$\\
\hline
Block size & $48 \times 46 \times 64$ & $102 \times 36 \times 36$ & $76 \times 32 \times 48$ & $64 \times 44 \times 64$ \\
\hline
Grid outer radius & $10^5 r_g$ & $2000 r_g$ & $150 r_g$ & $2000 r_g$ \\
\hline
Physical duration & $1.5 \times 10^5 r_g/c$ & $10^5 r_g/c$ & $10^4 r_g/c$ & $10^4 r_g/c$\\
\hline
Hardware computational cost & $3.8 \times 10^6$ GPU hours & $3.3 \times 10^4$ GPU hours & $18$ GPU hours & $1.1 \times 10^6$ GPU hours\\
\hline
System scale  & $5400$ V100 GPUs & $128$ V100 GPUs & $1$ V100 GPU & $6000$ V100 GPUs\\
\hline
Number of cells & $12-22 \times 10^9$ & $0.44 \times 10^9$ & $9.2 \times 10^6$ & $22 \times 10^9$\\
\hline
Number of real zone-cycles & $1.7 \times 10^{17}$  & $5.2 \times 10^{15}$ & $0.64 \times 10^{13}$ & $0.88 \times 10^{17}$\\
\hline
Number of effective zone-cycles & $1.5 \times 10^{18}$  & $1.8 \times 10^{16}$ & $1.6 \times 10^{13}$ & $4.4 \times 10^{17}$\\
\hline
Effective zone-cycles/s & $\gtrsim 1.1 \times 10^8$/GPU & $\gtrsim 1.5 \times 10^8$/GPU & $\gtrsim 2.5 \times 10^8$/GPU & $\gtrsim 1.1 \times 10^8$/GPU \\
\hline
LAT $\times$ GPU Speedup & 31 & 43 & 71 & 31\\
\hline
SMR Speedup ($ \# Timesteps$) & 3.3 & 1.42 & 1.17 & 6.7\\
\hline
AMR Speedup ($ \# Cells$) & 35 & 18 & 1 & 1\\
\hline
AMR Speedup ($ \# Timesteps$) & 53 & 20.7 & 1 & 1\\

\hline
Total Speedup & $1.9 \times 10^5$ & $2.3 \times 10^4$ & 83 & 208\\
\hline
\end{tabular}\\

\end{center}
\end{table*}

\section{Performance Summary}
In this article we have presented \hammer{}'s performance under a scenario of extreme complexity featuring an unusually high cell count, AMR, SMR, and $1000s$ of GPUs. In Table \ref{performance_table} we summarize \hammer{}'s performance spanning a range of complexity. These range from a typical $\sim 192^3$ GRMHD simulation (EHT-192) of a geometrically thick accretion disk that we ran on a single GPU during the GRMHD code comparison \citep{Porth2019} to the very thin disk (TD-4608) considered in this work. We also summarize \hammer{}'s performance in a mid-resolution GRMHD simulation of a thin accretion disk (TD-1728) with AMR on $128$ GPUs \citep{Liska2022}, and, in an $2304^3$ GRMHD simulation of a geometrically thick accretion disk (EHT-2304) without AMR on $6000$ GPUs \citep{Ripperda2022}.

In table \ref{performance_table} we calculate the speedup compared to \hammer{}'s performance on a $20$ core Intel Skylake CPU without any advanced features (AMR, SMR, LAT, GPUs). We assume that the timestep is not limited by the Courant condition in the $\varphi-$direction until a resolution of $N_{\varphi}=128$ due to cylindrification of the grid \citep{Tchekhovskoy2011}. A more detailed explanation of how these speed-up factors are calculated is given in the caption of table \ref{performance_table}. Under these assumptions, the attained speedup ranges from a factor $\sim 80$ for EHT-192 to a factor $\sim 1.9 \times 10^5$ for TD-4608, since TD-4608 benefits a lot from SMR and AMR. However, as the performance of EHT-2304 demonstrates, even in absence of any AMR, the speedup can exceed a factor $\gtrsim 200$ at higher resolution where squeezing of cells near the pole makes the simulation significantly more expensive without SMR (Sec.~\ref{ch1:sec:Pole}).

\subsection{Scalability}
\label{sec:weak-scaling}
The simulation presented in Sec.~\ref{ch1:sec:Scaling} scales up to $900$ OLCF Summit nodes and $5,\!400$ V100 GPUs. Each GPU on average contains 20 blocks, each of $48\times48\times64\sim50^3$ cells. The green dashed line in Fig.~\ref{ch1:fig:scaling} shows that with the use of SMR, AMR, and LAT, we obtain what might seem like a disappointing numerical efficiency (with the baseline set by the number of zone-cycles/s on a single GPU for a $150^3$ block) of $\sim 18\%$ and a weak scaling efficiency (using a single 6-GPU node as baseline with a grid containing $120$ blocks of size $48\times48\times64$) of $\sim 35\%$ at $900$ Summit nodes. One reason for the seemingly low parallel scaling efficiency is the difficulty of load-balancing that arrives with LAT: without LAT, the weak scaling efficiency increases to a respectable $\sim 60\%$. However, by decreasing the number of time steps, LAT speeds up the simulations by factors of $\sim4$ (on a single V100 GPU) to $\sim9$ (on 5400 V100 GPUs): this effectively brings the parallel numerical efficiency to $\sim 200\%$ at 900 Summit nodes = 5,400 V100 GPUs, as shown by the solid blue line in Fig.~\ref{ch1:fig:scaling}. This corresponds to (effectively) $\sim 1.175 \times 10^8$ zone cycles/s/GPU. Thus, the net result of LAT is a substantial increase in effective performance and parallel scaling efficiency.

The other reason for lower numerical efficiency in our fiducial problem compared to simpler test problems is the smaller block size. For illustration, consider an idealized weak scaling test, with a much larger block size of $150^3$ cells per GPU; we disable SMR, AMR, and LAT for this test. The orange dash-dotted line in Figure~\ref{ch1:fig:scaling} shows that this leads to an excellent parallel weak scaling efficiency of $\sim 80\%$ at 900 Summit nodes, primarily because inefficiencies in load balancing related to LAT go away (Sec.~\ref{ch1:sec:LAT}), and the MPI interconnect bandwidth is not anymore the main bottleneck due to the larger block size and, therefore, smaller fraction of boundary cells. Note that we would not be able to use such large blocks of $150^3$ cells for our fiducial problem as there would be an insufficient number of blocks per GPU to effectively load balance the grid and the blocks would be too big to properly focus on the very thin $h/r=0.02$ disk.

\section{Code Validation}
For the purpose of code development and validation, \hammer{} maintains an independently developed and maintained OpenMP and MPI parallelized C version running on CPUs. More specifically, we usually develop new features in the C version and, after extensive testing, port them to the CUDA/OpenCL version. Subsequently, we verify that the output of both versions agrees up to machine precision for various scenarios. We refer the reader to \citet{Porth2019} for code validation of \hammer{} against its peer codes, performed in the context of a community-wide GRMHD code comparison project and to the appendix of \citet{Liska2018A} for validation of \hammer{} for tilted accretion flows evolved on a spherical grid. We supplement these tests in this article with a suite of linear and non-linear wave tests, and two off-center spherical explosions.

\begin{figure}
\begin{center}
\includegraphics[width=\columnwidth,trim=0mm 0mm 0mm 0,clip]{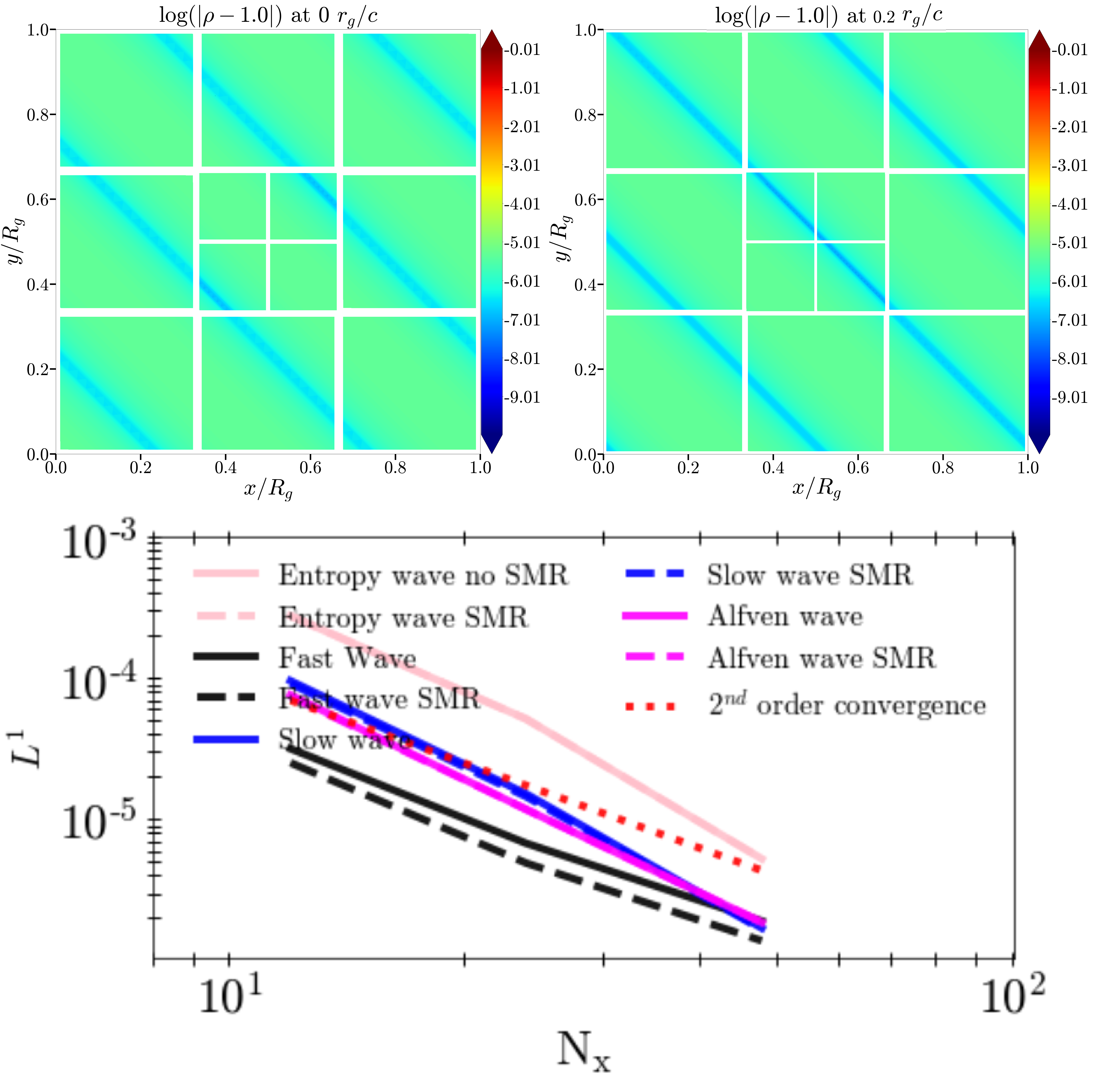}
\end{center}
\caption{\textbf{(Upper panels)} A density isocontour rendering at 2 different times for a fast wave. Block boundaries are illustrated using white lines. For our wave tests we start with a grid that has $3\times3\times3$ blocks and refine the innermost block. \textbf{(Lower panel)} A plot of the L1 error as function of the linear resolution $N_x$. \hammer{} achieves $2^{nd}$ order convergence for all waves irrespective of refinement boundaries. }
\label{fig:AMR_test}
\end{figure}

\subsection{Linear Wave Tests}
To verify $2^{nd}$ convergence of the boundary conditions between coarse and fine layers in \hammer{} we have performed a variety of 3D dimensional linear wave tests (with an $O(10^{-5})$) amplitude), which include the fast, slow, Alfven, and entropy waves (e.g. \citealt{Gammie2003}). We rotate the wavevector by 45 degrees with respect to the x, y and z axes on a Cartesian grid and force these waves to propagate through a grid refinement boundary, which also features a jump in timestepping (upper panels of Fig.~\ref{fig:AMR_test}). We evolve each respective wave for 2 periods and then compute the $L^1$ error, which is defined as the absolute value of the finite difference between the initial and final states for the primitive variables. In the lower panel of figure \ref{fig:AMR_test} we demonstrate that \hammer{}, as expected, converges at $2^{nd}$ order for all waves. 

\begin{figure}
\begin{center}
\includegraphics[width=1.0\columnwidth,trim=0mm 0mm 0mm 0,clip]{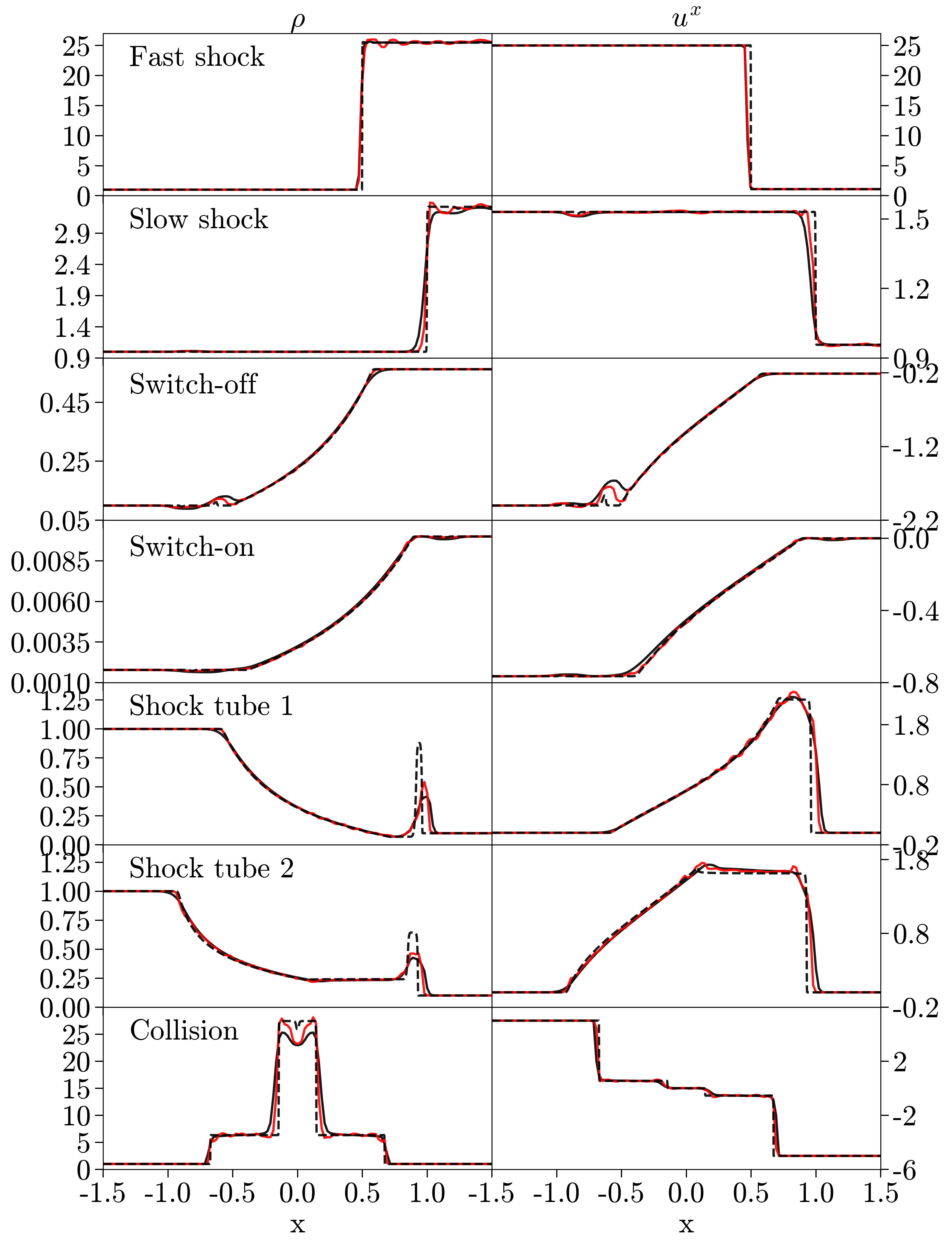}
\end{center}
\caption{Relativistic shock tubes taken from \citet{Komissarov1999} with density ($\rho$) in left panels and velocity ($u^x$) in right panels. Reconstruction with MINMOD limiter in black and with PPM limiter in red. The low resolution shock tubes with $\Delta x = 2.2 \times 10^{-2}$ (solid lines) are less sharp than the high resolution shock tubes with $\Delta x = 1.1 \times 10^{-3}$ (dashed lines), however they converge to the same solution. This verifies the ability of \hammer{} to handle shock waves in relativistic plasma.}
\label{fig:shocktube}
\end{figure}

\begin{figure*}
\begin{center}
\includegraphics[width=\linewidth,trim=0mm 0mm 0mm 0,clip]{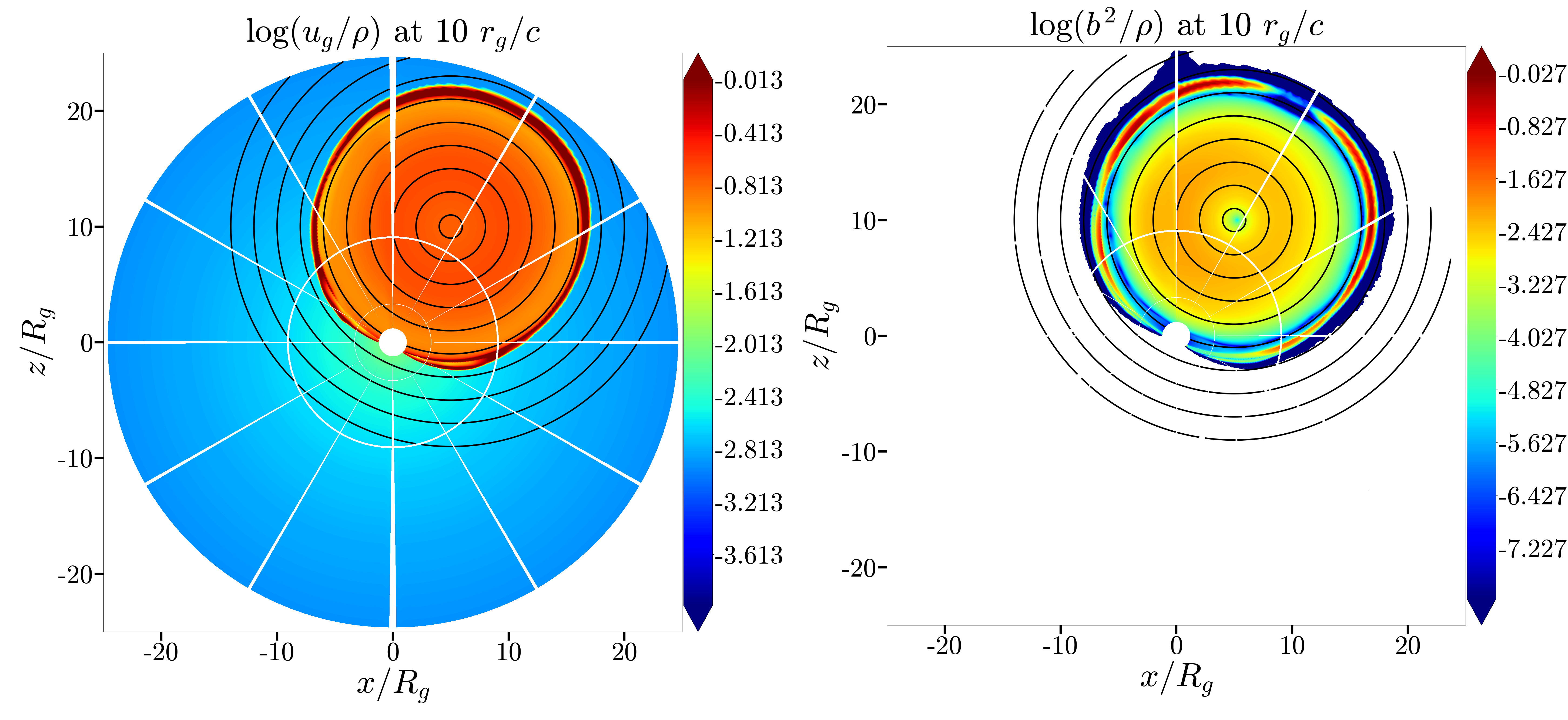}
\end{center}
\caption{\textbf{(Left panel)} The ratio between internal energy ($u_g$) and rest mass density ($\rho)$ during a spherical off-center explosion that passes through 2 internal and 1 external SMR layer in addition to the polar axis. The different meshblocks are seperated by white lines. Black lines illustrate the distance from the center of the explosion in steps of $1 r_g$. The front of the shockwave becomes less resolved at larger radii due to the logarithmic spacing of the grid. \textbf{(Right panel)} The ratio between the magnetic energy ($b^2$) and rest mass density ($\rho)$ during a magnetized off center explosion that passes through 2 internal and 1 external SMR layer in addition to the polar axis. When the explosion passes through the polar singularity only minor artifacts in the internal energy and magnetic field strength are detected, which is consistent with other (GR)MHD codes that utilize similar (transmissive) boundary conditions (e.g. \citealt{Mewes2018, Mewes2020}).}
\label{fig:blastwave}
\end{figure*}

\subsection{Non-Linear Wave Tests}
To verify the ability of \hammer{} to capture shock waves under challenging conditions we perform a series of relativistic shock tube tests as described in \citet{Komissarov1999}. We evolve the shock tubes on a 1D Cartesian grid in Minkowskian space with no refinement and consider both a MINMOD and PPM slope limiter (without contact steepener or shock flattener). We use a Courant factor $C_f=0.8$ except for the fast shock for which we use $C_f=0.2$. We compare in Figure \ref{fig:shocktube} an extremely high resolution shock tube ($\Delta x = 1.1 \times 10^{-3}$) to a lower resolution shock tube ($\Delta x = 2.2 \times 10^{-2}$). The shock fronts in the low resolution shock tube are, as expected, less resolved. While the very diffusive MINMOD limiter does not produce spurious oscillations near shock fronts, some relatively small spurious oscillations are detectable with the less diffusive PPM limiter. This is a well known disadvantage of PPM limiters that can be addressed with an optional shock flattener implemented in \hammer{} which reduces the order of reconstruction near shock fronts \citep{ColellaWoodward1984}. These tests verify the robustness and stability of the shock capturing scheme in \hammer{} when handling shocks in relativistic plasma.

\subsection{Off-center Spherical Explosion}
To quantify numerical errors associated with gas passing through the polar singularity and internal/external SMR layers, we perform an off-center explosion that propagates through the poles. During this test we use 1 layer of external SMR, 2 layers of internal SMR, and 4 levels of local adaptive timestepping. The test is performed on a logarithmically spaced spherical grid instead of a Cartesian grid to also validate the correct implementation of of the geometric source terms. The total resolution is $300 \times 240 \times 240$. The initial pressure and density are set to a constant $p=3 \times 10^{-5}$ and $\rho = 10^{-4}$ within a sphere of radius $r_s=2 r_g$ that is centered at $(x_0,y_0, z_0)=(3, 0, 10)r_g$. We use an exponential cutoff to make the transition between the sphere and ambient medium smoother. In addition to the hydrodynamic explosion, we perform a similar test where the explosion is threaded by a magnetic field described by vector potential $A_{\phi} \propto (r_s -\sqrt{(x-x_0)^2+(y-y_0)^2+(z-z_0)^2})$ for $r < r_s$ and $A_{\phi}=0$ for $r > r_s$. The magnetic field is subsequently normalized such that $\beta^{max}=p_{g}^{max}/p_{b}^{max}=10$. Here $p_{g}^{max}$ and $p_{b}^{max}$ are the maximum gas and magnetic pressure in the grid.

The results of these off-center explosions are illustrated in the left (hydrodynamic explosion) and right (magnetized explosion) panels of Figure \ref{fig:blastwave}. Both explosions manage to maintain their circular shape when passing through the polar axis. The minor distortions around the polar singularity are similar to those observed in other GRMHD codes that use transmissive boundary conditions (e.g \citealt{Mewes2018, Mewes2020}). This demonstrates that \hammer{} is capable of accurately evolving plasma around the pole and can even maintain some form of convergence when the gas passes directly through the polar singularity.

To avoid repeated passage of, for example, a kink-unstable jet through the polar singularity \hammer{} has the ability to `rotate' the setup with respect to the spherical grid such that the jet propagates along the equatorial plane instead (e.g. \citealt{Gottlieb2021, Gottlieb2022A, Gottlieb2022B}). The physical equivalence between rotated and non-rotated setups was demonstrated in the appendix of \citet{Liska2018A}.

\section{Discussion}
\label{ch1:sec:Discussion}
\subsection{Advancements in Computational Methods}
\label{sec:advanc-comp-meth}
The extensive optimizations presented in this paper bring state of the art GRMHD simulations to the exascale level, e.g. to clusters with more than $10^{18}$ FLOP/s of FP64 performance, using a highly efficient grid capable of tackling problems with many orders of magnitude in scale separation. We carried out scaling tests up to $5400$ NVIDIA V100 GPUs on OLCF Summit, culminating in our fiducial simulation: the largest GRMHD simulation to date that utilizes $4$ AMR levels, with the block size of $48\times64\times64$ cells, and achieves an astonishingly high effective resolution in the disk of $13440 \times 4608 \times 8096$ cells. The simulation cost is around 4 million NVIDIA V100 GPU-hours, which corresponds to roughly 800 million Sky Lake CPU core hours (Sec.~\ref{ch1:sec:Benchmarks}).

\hammer{} features a set of optimizations that collectively boost the performance of GRMHD simulations by at least 2 orders of magnitude for simple problems that do not benefit from an adaptive mesh. This increases to 5 orders of magnitude for problems requiring adaptive meshes and very high resolutions in the polar region (see Sec.~\ref{sec:numerical-grid}). This speedup is attained by the following means. An NVIDIA V100 GPU typically provides a factor of $\sim 40$ speedup compared to a non AVX-vectorized version of \hammer{} running on 16 Intel Skylake cores (AVX vectorization improves the performance by a factor $\sim 3$). Large gains also come from the advanced treatment of the polar region during SMR/AMR which provides a typical factor $5{-}20$ reduction in the number of timesteps, depending on the resolution. Local adaptive timestepping reduces the number of timesteps by another factor of $\sim 3{-}10$ depending on the resolution of the grid. Finally, AMR can reduce the number of cells in the grid by a factor of $\sim 10{-}100$ for many complex problems such as thin accretion disks and large-scale jets.

This paves the way for a next generation of GRMHD simulations with unprecedented resolutions and record-breaking simulated timescales, substantially advancing the understanding of accreting black holes. For example, \hammer{} made it possible to simulate even the most challenging accretion disk configurations naturally expected in nature (i.e., thin and tilted) over timescales that are relevant to astronomical observations \citep{Liska2018C,Liska2019A,Liska2019b}. \citet{Liska2018C} demonstrated, for the first time in GRMHD, that tilted thin magnetized disks align with the black hole midplane as theoretically predicted \citep{bp75}, solving a 40 year old problem in astrophysical disk dynamics. Further, not only this article, but also \citet{Liska2019b} and \citet{Musoke2022}, simulated disk tearing in highly tilted thin disks (Fig.~\ref{ch1:fig:AMR}), potentially explaining some types of QPOs observed in XRBs (e.g., \cite{Kalamkar2016}), as well as holding important consequences for galactic evolution (e.g., \cite{King2005}). Using \hammer{}, \citet{Chatterjee2019} simulated the longest extent (in both temporal and spatial dimensions) relativistic jets from an accreting black hole in 2D GRMHD, making it possible to follow the jet over a range of $\gtrsim 5$ orders of magnitude in distance and enable the study of the jet's deceleration process as it carves out its path through the environment. Following this line of study, \hammer{} would finally make it possible to simulate black hole-launched jets and connect distances from the event horizon up to extra-galactic distances, closing the gap between GRMHD and cosmological simulations of galaxy evolution. 

\hammer{} also played an important role in interpreting (e.g. \citealt{Porth2019, Chatterjee2020, Ripperda2022, EHT_SGR_V}) the black hole images from Event Horizon Telescope (EHT) observations of the supermassive black holes M87$^*$ (e.g. \citealt{EHT_M87_1, EHT_M87_V, EHT_M87_8}) and Sagittarius~A$^*$ (e.g. \citealt{EHT_SGR_1, EHT_SGR_IV}). In particular, \hammer{} enabled the first study of how general relativity warps jets (and their accretion disks) that are misaligned with the BH spin axis, ultimately affecting the near-event horizon image of a black hole \citep{Liska2018C, Chatterjee2020}. The ability to carry out the simulations at extremely high resolutions enables \hammer{} to attain in global simulations \citep{Ripperda2022} the resolutions typically associated with local simulations. Such global modeling has already enabled a significant improvement in our understanding of how black holes consume and eject gas. It has been thought that the formation of relativistic jets by accreting black hole requires the presence of large-scale vertical magnetic flux to begin with, and that absent such field -- e.g., if only purely toroidal magnetic field is present -- the simulations do not produce jets \citep{bhk08,Mckinney2012}. However, our simulations have revealed that such configurations are actually capable of generating large-scale vertical magnetic flux and forming relativistic jets \citep{Liska2018B}.

Summing up, \hammer{} is in a prime position to address important questions about the dynamic nature of accretion disks and jets such as the hitherto unexplained trigger behind the observed change in disk geometry from thick to thin (e.g., \citealt{Belloni2010, Liska2022}), the evolution of jet morphologies (known as the FR-I/II dichotomy \citealt{fr74}), the origins of particle acceleration in both disks and jets \citep{sironi2015, Ripperda2022}. The high speed of \hammer{} makes it attractive to account for departures from ideal MHD physics such as radiation (e.g. \citealt{Scadowski2013, Ryan2017, Foucart2018}) and synchrotron cooling of electrons (e.g., \citealt{Fragile2009A,Dibi:12,Drappeau:13, Yoon2020}), resistivity (e.g. \citealt{Ripperda_BHAC,ripper2019, Mignone2019}) and two-temperature thermodynamics (e.g. \citealt{Ressler2015, Chael2017, Ryan2018, Liska2022}), which may lead to important insights into many of the above mentioned problems. Currently, \hammer{}'s speed is only (MPI-bandwidth) limited by waiting times for inter-node communication, which leaves time for extra computations. Therefore, more complex physics can be implemented (almost) free of charge. Such numerical experiments are expected to scale well on exascale systems, even if those systems feature slower inter-node communication than in the simulations presented here and conducted at OLCF Summit.

\begin{figure*}
\begin{center}
\includegraphics[width=0.8\linewidth,trim=0mm 0mm 0mm 0,clip]{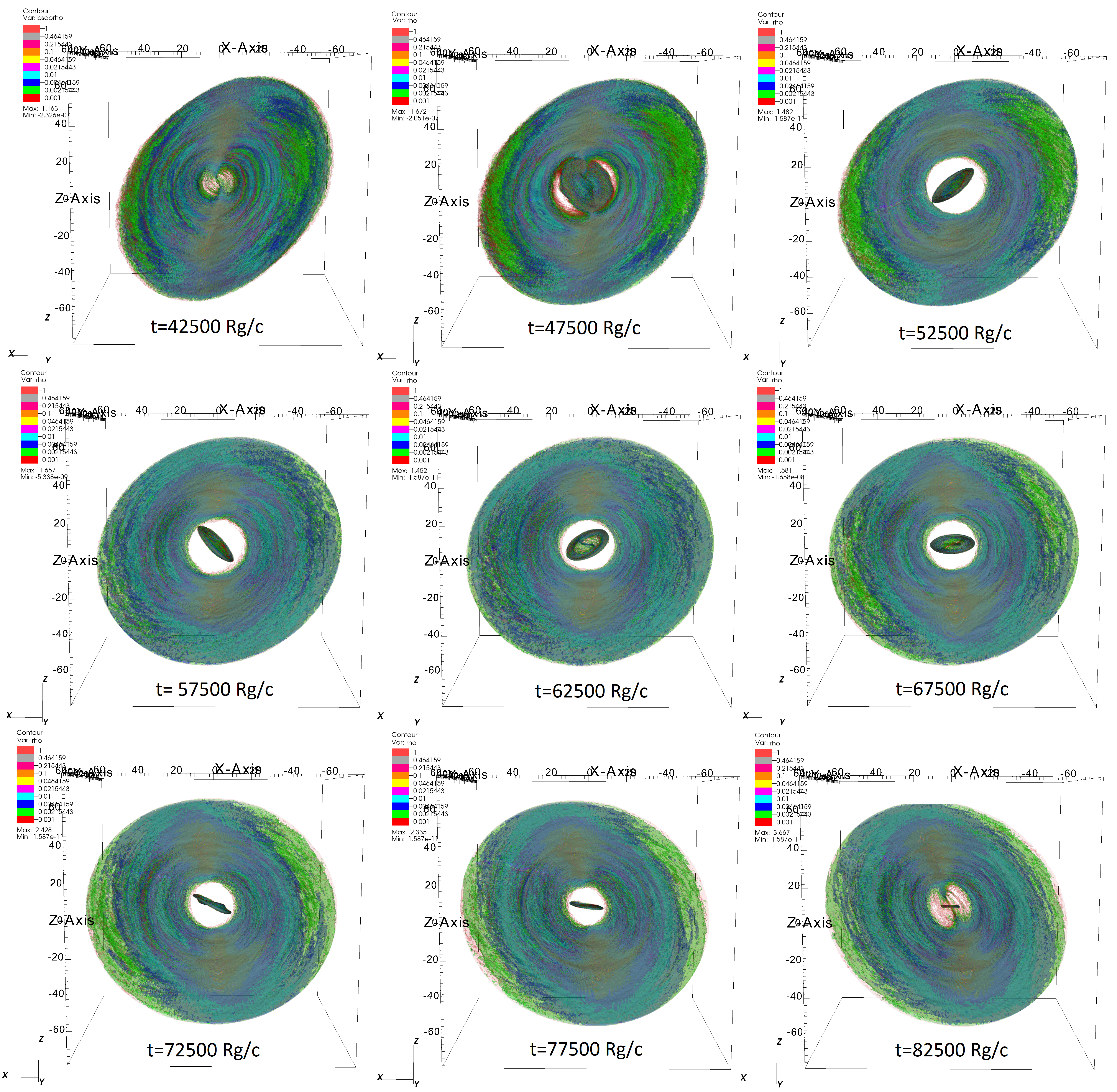}
\end{center}
\caption{A 3D isocontour rendering of density at different times in our simulation, showing a single tearing cycle of the disk: at the initial time shown, $t = 42,500r_g/c$ (top left panel), the disk is intact. However, at $t = 47,500r_g/c$, the inner disk of size $\sim 20r_g$ tears off from the outer disk and starts to precess independently of the outer disk. Eventually, the inner precessing disk gets consumed by the black hole, and the entire disk appears intact again at the last time shown, $t = 82,500r_g/c$ (bottom right panel). During each tearing cycle the torque from the spinning black hole overcomes the viscous torques holding the disk together and the disk tears apart. This results in a rapidly precessing inner subdisk physically detached from the outer disk. This subdisk can sustain several precession periods before it shrinks and disappears into the black hole and the tearing cycle repeats. Disk tearing and precession is an attractive model to explain Type-C quasi periodic oscillations in the lightcurves of XRBs (e.g. \citealt{Klis2006}).}
\label{ch1:fig:volumeplot}
\end{figure*}

\subsection{Scientific Results}
\label{sec:scientific-results}
For the scaling tests presented in this article we have used a real-world problem involving a highly tilted (by $65^{\circ}$) very thin ($h/r=0.02$) accretion disk threaded by a strong toroidal magnetic field ($\beta \sim 10$) around a rapidly spinning black hole ($a\sim0.94$). These simulations are more extensively described in \citet{Musoke2022}. Interestingly, unlike thick disks threaded with toroidal magnetic field that tend to develop large-scale poloidal magnetic flux through a dynamo process and associated relativistic jets \citep{Liska2018B}, our thin disk did not produce any jets, as seen in Fig.~\ref{ch1:fig:AMR}. In fact, due to the complete absence of jets, the outcome resembles the high/soft state in XRBs, and as such this is the first GRMHD simulation to reach this exceptionally challenging jet-less regime, in the complete absence of initial poloidal magnetic flux. While in previous work we needed to seed the accretion disk with a rather strong poloidal magnetic field as is only expected in the intermediate states \citep{Liska2018C, Liska2019A,Liska2019b}, the powerful NVIDIA V100 GPUs of OLCF Summit allowed us to resolve the weak toroidal magnetic field with $\gtrsim 20$ cells per MRI wavelength and the disk scale height with $25-30$ cells.

As illustrated in Figs.~\ref{ch1:fig:AMR} and \ref{ch1:fig:volumeplot} and in the accompanying \href{https://www.youtube.com/watch?v=rIOjKUfzcvI}{movie}, we find that tilted thin accretion disks, initially threaded with a purely toroidal magnetic field (i.e., containing no poloidal magnetic flux to begin with), can tear apart into multiple precessing sub-disks. We also find that, even in the absence of a \citet{bz77} jet, the inner regions of a magnetized accretion disk can sporadically align with the black hole midplane, as predicted more than 4 decades ago by \citet{bp75} and seen in such disks for the first time (see Fig.~\ref{ch1:fig:AMR}). The \href{https://www.youtube.com/watch?v=rIOjKUfzcvI}{movie} of the simulation shows multiple ($\gtrsim4$) cycles of such tearing events that result in the formation of a precessing inner sub-disk of about the same size. The emission from such precessing sub-disks is expected to be quasi-periodic. Thus, the tearing of tilted disks is an attractive mechanism for producing coherent quasi-periodic oscillations (QPOs), now testable via GRMHD simulations for the first time (e.g. \citealt{Musoke2022}). Improved understanding of the evolution of QPO frequency and amplitude during XRB outbursts may lead to unique constraints on black hole spin magnitude and/or disk geometry (e.g. \citealt{Ingram2009, Motta2015}) and provide new insights into the physics driving XRB spectral state transitions (e.g. \citealt{Ferreira2006, Begelman2014, Marcel2018A, Marcel2018B, Liska2018B,  Liska2018C}).

\section{Acknowledgments}
\label{ch1:sec:Ack}
An award of computer time was provided by the Innovative and Novel Computational Impact on Theory and Experiment (INCITE) and the ASCR Leadership Computing Challange (ALCC) programs under award PHY129. This research used resources of the Oak Ridge Leadership Computing Facility, which is a DOE Office of Science User Facility supported under Contract DE-AC05-00OR22725 using NERSC award ALCC-ERCAP0022634. We acknowledge PRACE for awarding us access to JUWELS Booster at GCS@JSC, Germany. ML was supported by the John Harvard Distinguished Science and ITC Fellowships, and the 21-ATP21-0077 NASA ATP award. MK was supported by the Netherlands Organisation for Scientific Research (NWO) Spinoza Prize, CH by the Amsterdam Science Talent Scholarship, AI by a Royal Society University Research Fellowship, AT by the National Science Foundation grants AST-2206471, AST-2009884,AST-2107839, AST-1815304, OAC-2031997 and AST-1911080, and SM, KC and DY by the NWO VICI grant (no.~639.043.513).

\bibliography{new.bib}{}
\bibliographystyle{aasjournal}



\end{document}